\documentclass[aps,prb,twocolumn,floats,superscriptaddress,tighten,letterpaper,floatfix]{revtex4-2}
\usepackage{bbm,mathtools}
\usepackage[utf8]{inputenc}
\usepackage{amssymb}
\usepackage{amsbsy}
\usepackage{amsmath}
\usepackage{graphicx}
\usepackage{graphics}
\usepackage{setspace}
\usepackage{array}
\usepackage{color}
\usepackage{xcolor}
\usepackage{fontenc}
\usepackage{textcomp}
\usepackage{rotating}
\usepackage{bm}
\usepackage{braket}
\usepackage[colorlinks=true,linkcolor=black,bookmarksopen=false,urlcolor=blue,citecolor=blue]{hyperref}
\hypersetup{pdfpagemode=UseNone}
\usepackage{chngcntr}
\usepackage{subfigure}

\usepackage{comment}
\usepackage{color}
\usepackage{slashed}
\usepackage[colorlinks=true,linkcolor=black,bookmarksopen=false,urlcolor=blue,citecolor=blue]{hyperref}
\usepackage{amsmath}
\usepackage{enumitem}
\usepackage{revsymb}

\newcommand{\calA}{{\cal A}}
\newcommand{\calC}{{\cal C}}
\newcommand{\calD}{{\cal D}}

\newcommand{\calS}{{\cal S}}

\newcommand{\hThe}{\hat{\Theta}}
\newcommand{\hSig}{\hat{\Sigma}}
\newcommand{\hsig}{\hat{\sigma}}
\newcommand{\hs}{\hat{s}}

\newcommand{\hD}{\hat{D}}

\newcommand{\hq}{\hat{q}}
\newcommand{\hW}{\hat{W}}
\newcommand{\hR}{\hat{R}}
\newcommand{\hX}{\hat{X}}
\newcommand{\Xdag}{{X}^{\dagger}}
\newcommand{\hXdag}{\hat{X}^{\dagger}}
\newcommand{\hY}{\hat{Y}}
\newcommand{\Ydag}{{Y}^{\dagger}}
\newcommand{\hYdag}{\hat{Y}^{\dagger}}
\newcommand{\hEsig}{\hat{E}^{\sigma}}

\newcommand{\hWdag}{\hat{W}^{\dagger}}

\newcommand{\hqsp}{\hat{q}_{\mathsf{sp}}}

\newcommand{\homega}{\hat{\omega}}
\newcommand{\hM}{\hat{M}}
\newcommand{\hphi}{\hat{\phi}}
\newcommand{\hphicl}{\hat{\phi}^{\sfcl}}
\newcommand{\hphiq}{\hat{\phi}^{\sfq}}

\newcommand{\Idag}{I^{\dagger}}
\newcommand{\Jdag}{J^{\dagger}}

\newcommand{\frakD}{\mathfrak{D}}

\newcommand{\sfb}{\mathsf{b}}
\newcommand{\sfT}{\mathsf{T}}

\newcommand{\sfcl}{\mathsf{cl}}
\newcommand{\sfq}{\mathsf{q}}

\newcommand{\sfMFL}{\mathsf{MFL}}

\newcommand{\sfmax}{\mathsf{max}}
\newcommand{\sfeff}{\mathsf{eff}}

\newcommand{\sfdc}{\mathsf{dc}}

\newcommand{\sfscr}{\mathsf{scr}}

\newcommand{\sfintI}{\mathsf{int\,I}}
\newcommand{\sfintII}{\mathsf{int\,II}}
\newcommand{\sfsemi}{\mathsf{semi}}

\newcommand{\Deltacl}{\Delta^{\sfcl}}
\newcommand{\Deltabarcl}{\bar{\Delta}^{\sfcl}}
\newcommand{\Deltaq}{\Delta^{\sfq}}
\newcommand{\Deltabarq}{\bar{\Delta}^{\sfq}}
\newcommand{\phicl}{\phi^{\sfcl}}
\newcommand{\phiq}{\phi^{\sfq}}

\newcommand{\veps}{\varepsilon}

\newcommand{\vphi}{\varphi}
\newcommand{\gbar}{\bar{g}}

\newcommand{\htau}{\hat{\tau}}

\newcommand{\gammael}{\gamma_{\mathsf{el}}}

\newcommand{\mb}{m_{\mathsf{b}}}

\newcommand{\boson}{\mathsf{b}}
\newcommand{\fermion}{\mathsf{f}}
\newcommand{\AMFL}{{\cal A}_{\mathsf{MFL}}}
\newcommand{\mfl}{\mathsf{MFL}}

\newcommand{\diag}{\mathsf{diag}}
\newcommand{\Tr}{\mathsf{Tr}\, }

\newcommand{\im}{\,\text{Im}\,}
\newcommand{\re}{\,\text{Re}\,}

\newcommand{\e}{\text{e}}

\newcommand{\vex}[1]{\bm{\mathrm{#1}}}
\newcommand{\Nabla}{\bm{\nabla}}

\begin{document}

\title{
Enhancement of Superconductivity in a Dirty Marginal Fermi Liquid
}

\author{Tsz Chun Wu}
\affiliation{Department of Physics and Astronomy, Rice University, Houston, Texas 77005, USA}
\author{Patrick A. Lee}
\affiliation{Department of Physics, Massachusetts Institute of Technology, Cambridge, Massachusetts 02139, USA}
\author{Matthew S. Foster}
\affiliation{Department of Physics and Astronomy, Rice University, Houston, Texas 77005, USA}
\affiliation{Rice Center for Quantum Materials, Rice University, Houston, Texas 77005, USA}

\date{\today}

\begin{abstract}
We study superconductivity in a two-dimensional, disordered marginal Fermi liquid. At the semiclassical level, the transition temperature $T_c$ 
is strongly suppressed because marginal Fermi liquid effects destroy well-defined quasiparticles. However, we show that interference between 
quantum-critical collective modes must be included, and these \emph{enhance} $T_c$, violating Anderson's theorem. 
Our results suggest that phase coherence in a disordered, quantum-critical system can survive and manifest in the form of collective excitations, despite the absence of coherent quasiparticles. 
\end{abstract}

\maketitle

\section{Introduction}

Strange-metal behavior has been a persistent mystery for over 30 years \cite{MFL_Varma_CuO_PRL_89} 
and by now has been observed in a wide range of quantum materials including the cuprates \cite{NFL_CuO_review_Kivelson_Nat_15} 
and 
twisted bilayer graphene \cite{TBG_Cao_Nat_2018}. 
Recent theoretical results suggest that 
both
quantum-critical interactions \textit{and} 
quenched
disorder are key ingredients for strange metallicity 
\cite{SYK_Patel_PRB_21,SYK_Patel_PRB_2022,SYK_Guo_Patel_linearT_PRB_2022,SYK_review_Sachdev_RMP_2022}.
In particular, Refs.~\cite{SYK_Patel_linearT_arxiv_2022,MFL_Wu_Liao_Foster_PRB_22} 
have proposed microscopic models based on the interplay between a critical mode and 
impurity scattering
that address aspects of the 
disordered marginal Fermi liquid (MFL) phenomenology. 
The next challenge is to understand the interplay of critical interactions and disorder on superconductivity \cite{NFL_SC_Kivelson_PNAS_2017,NFL_SC_Chubukov_I_PRB_20,NFL_SC_Chubukov_II_PRB_20,NFL_SC_Chubukov_III_PRB_20,NFL_SC_Chubukov_IV_PRB_21,
NFL_SC_Chubukov_V_PRB_21,NFL_SC_Chubukov_VI_PRB_21,NFL_SC_Kamenev_PRL_2022,NFL_SC_SYK_Sachdev_arxiv_2022,disO_SC_MFC_Sri_Burmistrov_arxiv_2022,Setty_PRB_2022}.
In this paper we address a fundamental paradox: 
Why does superconductivity, a macroscopic, coherent many-body quantum state, arise at anomalously high temperatures from a collision-dominated, seemingly incoherent and dirty strange metal?

We extend Ref.~\cite{MFL_Wu_Liao_Foster_PRB_22} to include pairing due to an attractive interaction $W$. 
The phenomenological parameter $W$ can originate from phonons or from integrating out short-scale fluctuations of the critical mode.  
Despite the \emph{absence of well-defined quasiparticles} due to Planckian dissipation, 
we predict a strong tendency towards superconductivity in a disordered MFL, with a transition temperature
\begin{equation}
\label{eq:Tc}
	T_c 
	\sim 
	\begin{cases}
		 g^2 (\nu_0 W_{\sfeff}) \, {\calC} / \sigma_{\sfdc},
		\qquad
		& 
		T_c \gtrsim T_*,
	\\
		 g^2 (\nu_0 W_{\sfeff})^2 / \sigma_{\sfdc},
		\qquad
		& 
		T_c \ll T_*.
	\end{cases}
\end{equation}
In Eq.~(\ref{eq:Tc}), 
$g$ is the Yukawa coupling between fermions and quantum-relaxational (critical) bosons,
$\sigma_{\sfdc}$ 
is the total dc conductivity 
of the non-superconducting MFL state in units of $e^2/\hbar$, 
$\nu_0$ is the density of states (DoS) per channel,  
and 
$W_{\sfeff}$ is the effective BCS attractive interaction strength renormalized by MFL effects.
The dimensionless constant $\calC$ is non-universal, depending upon the model parameters.
While Ref.~\cite{MFL_Wu_Liao_Foster_PRB_22} focused on the regime where the bosons are characterized by a 
thermal mass, Eq.~(\ref{eq:Tc}) also includes the low-$T$ limit $T < T_*$, where $ T_*$ marks the crossover to the regime
dominated by dynamical screening \cite{disO_SC_MFC_Sri_Burmistrov_arxiv_2022}.

Eq.~(\ref{eq:Tc}) gives a non-BCS prediction for $T_c$, proportional to a power of the interaction coupling $W_{\sfeff}$. 
The mechanism leading to this result is a \emph{quantum interference} (Maekawa-Fukuyama \cite{disO_SC_Fukuyama_PhysicaB_1982}) process,
mediated by the critical boson mode. This process is of the same nature as that responsible for Anderson localization, 
and implies that a dirtier normal state (smaller $\sigma_{\sfdc}$) yields a \emph{larger} $T_c$. 
Eq.~(\ref{eq:Tc}) should be compared to an artificial semiclassical limit that neglects interference,
wherein we find a strongly suppressed transition temperature,
\begin{align}\label{TcS}
	T_c^{\mathcal{S}}
	\simeq
	\Lambda
	\,
	\exp\left[
		-
		\frac{1}{\gbar^2}
		\left(
			e^{\gbar^2 / \nu_0 W} - 1
		\right)
	\right].
\end{align}
Here $\Lambda$ is a UV cutoff and $\gbar^2 \sim g^2/\gammael$ (dimensionless), where $\gammael$ is the semiclassical elastic 
impurity scattering rate. In the limit of vanishing fermion-boson coupling $\gbar \rightarrow 0$, Eq.~(\ref{TcS}) 
reduces to the BCS result $T_c^{\mathsf{BCS}} \propto e^{-1 / (\nu_0 W)}$. Finite $\gbar$ exponentially suppresses the transition due to the lack of well-defined quasiparticles in the MFL. For small $\nu_0 W \sim \nu_0 W_{\sfeff} \ll 1$ and order one $\gbar$ and $\sigma_{\sfdc}$, we can have 
$T_c^{\mathcal{S}} \ll T_c^{\mathsf{BCS}} \ll T_c$. This indicates the crucial role played by quantum interference. 
To put our results in context, we next briefly review analogous physics in conventional disordered superconductors
\cite{
disO_SC_AndersonThm_Anderson_1959,
disO_SC_AndersonThm_Gorkov_JETP_1959,
disO_SC_Fukuyama_PhysicaB_1982,
disO_SC_PALee_PRB_1985,
disO_review_PALee_RMP_1985,
disO_SC_Finkelstein_Fukuyama_PhysicaB_1994,
disO_review_Kirkpatrick_94,
disO_SC_Feigelman_PRB_2000,
disO_SC_Num_Trivedi_PRL_1998,
disO_SC_Num_Trivedi_PRB_2001,
disO_SC_Num_Garcia-Garcia_PRB_2020,
disO_SC_MFC_Feigelman_PRL_2007,
disO_SC_MFC_Feigelman_AnnPhy_2010,
disO_SC_MFC_Burmistrov_PRL_2012,
disO_SC_MFC_Foster_PRL_2012,
disO_SC_MFC_Foster_PRB_2014,
disO_SC_MFC_Garcia-Garcia_PRB_2015,
disO_SC_MFC_Burmistrov_PRB_2015,
disO_SC_Konig_Mirlin_PRB_2015,
disO_SC_MFC_Burmistrov_PRR_2021,
disO_SC_MFC_Burmistrov_AnnPhys_2021,
disO_SC_random_inhomo_Kivelson_PRB_2004,
disO_SC_random_inhomo_Kivelson_PRB_2005,
disO_SC_random_inhomo_Scalettar_PRB_2006,
disO_SC_random_inhomo_Scalettar_PRB_2007,
disO_SC_random_inhomo_Kivelson_PRB_2007,
disO_SC_random_inhomo_Scalettar_PRB_2008,
disO_SC_random_inhomo_Kivelson_PRB_2008,
disO_SC_random_inhomo_Refael_PRB_2008,
disO_SC_random_inhomo_Garcia-Garcia_PRA_2010,
disO_SC_random_inhomo_Kravtsov_2012,
disO_SC_random_inhomo_Scalettar_PRB_2012,
disO_SC_random_inhomo_DellAnna_PRB_2013,
disO_SC_random_inhomo_Andersen_PRB_2018,
disO_SC_random_inhomo_Kivelson_PRB_2018,
disO_SC_Feigelman_Nat_20,
disO_SC_Finkelstein_PRB_2022,
disO_SC_MFC_Sri_Burmistrov_arxiv_2022,
disO_SC_MFC_Garcia-Garcia_PRB_2020,
disO_SC_MFC_Zhang_Foster_PRB_2022},
see also Secs.~\ref{sec:BCSMFC} and \ref{sec:MFCDirtMix}.

Anderson's theorem dictates that $s$-wave superconductivity is immune to non-magnetic disorder as long as the normal state is a good metal \cite{disO_SC_AndersonThm_Anderson_1959,disO_SC_AndersonThm_Gorkov_JETP_1959}. 
However, a seminal work by Maekawa and Fukuyama revealed that Coulomb interactions can suppress superconductivity at the quantum level, 
due to Anderson localization \cite{disO_SC_Fukuyama_PhysicaB_1982}.
The latter serves as a precursor to the superconductor-insulator transition 
\cite{disO_SC_Finkelstein_Fukuyama_PhysicaB_1994,disO_SC_MFC_Burmistrov_AnnPhys_2021}.

Interestingly, recent developments suggest a new twist of this theme: short-ranged (e.g., externally screened Coulomb) interactions 
and random or structural inhomogeneity can enhance superconductivity near the Anderson metal-insulator transition \cite{
disO_SC_MFC_Feigelman_PRL_2007,
disO_SC_MFC_Feigelman_AnnPhy_2010,
disO_SC_MFC_Burmistrov_PRL_2012,
disO_SC_MFC_Foster_PRL_2012,
disO_SC_MFC_Foster_PRB_2014,
disO_SC_MFC_Burmistrov_PRB_2015,
disO_SC_MFC_Garcia-Garcia_PRB_2015,
disO_SC_MFC_Burmistrov_AnnPhys_2021,
disO_SC_MFC_Burmistrov_PRR_2021,
disO_SC_MFC_Garcia-Garcia_PRB_2020,
disO_SC_MFC_Zhang_Foster_PRB_2022,
disO_SC_MFC_Sri_Burmistrov_arxiv_2022}. 
Owing to quantum interference, single-particle wavefunctions exhibit multifractality and strong state-to-state overlaps \cite{disO_SC_abrahams1981}, 
resulting in larger interaction matrix elements that promote the pairing amplitude \cite{disO_SC_MFC_Feigelman_PRL_2007,disO_SC_MFC_Feigelman_AnnPhy_2010}.   
The detailed mechanism 
\cite{
disO_SC_MFC_Feigelman_PRL_2007,
disO_SC_MFC_Feigelman_AnnPhy_2010,
disO_SC_MFC_Burmistrov_PRL_2012,
disO_SC_MFC_Foster_PRL_2012,
disO_SC_MFC_Burmistrov_PRB_2015}
involves mixing between the Cooper and other interaction channels,
and is in fact identical to the Maekawa-Fukuyama process 
\cite{
disO_SC_Finkelstein_Fukuyama_PhysicaB_1994,
disO_SC_Fukuyama_PhysicaB_1982} 
that suppresses $T_c$ when Coulomb interactions are included. 
(See Sec.~\ref{sec:MFCDirtMix} for a review of why Coulomb interactions suppress $T_c$ instead of enhancing it.)

We note that the superconducting transition temperature is determined by the \emph{minimum} of the pairing amplitude and the superfluid stiffness \cite{Scalapino1993,Trivedi1996}; multifractality can enhance the amplitude, while the stiffness is typically monotonically suppressed by increasing inhomogeneity. Close (but not too close) to the single-particle localization transition, the inhomogeneous amplitude itself spans a finite portion of the sample and is not fractal \cite{disO_SC_MFC_Feigelman_AnnPhy_2010}. This regime is far from the superconducting-island scenario \cite{disO_SC_Num_Trivedi_PRB_2001}, and the stiffness exceeds the average fractality-enhanced amplitude. The transition temperature should then be determined by the latter \cite{disO_SC_MFC_Zhang_Foster_PRB_2022}.

Quantum 
effects like multifractality might be expected to play no role in a strongly correlated non-Fermi liquid lacking long-lived quasiparticles. 
An intensively studied paradigm for such systems consists of two-dimensional (2D) fermions at finite density coupled to quantum-critical bosons \cite{
NFL_SSLee_Review_18,
NFL_Subir_book_CUP_2011,
NFL_QMC_review_Berg_AnnRevCMP_19,
SYK_review_Sachdev_RMP_2022}. 
Due to the large decay rate, the Landau quasiparticle paradigm breaks down in these theories.

In a dirty quantum-critical system, impurities can dramatically modify the self-energies of the bosons and fermions, due to disorder smearing. 
At finite temperature $T$ 
and at the quantum critical point, 
the (retarded) bosonic propagator acquires a quantum relaxational form 
\cite{
SYK_Patel_PRB_21,
SYK_Patel_linearT_arxiv_2022,
MFL_Wu_Liao_Foster_PRB_22,
NFL_Subir_book_CUP_2011}
\begin{equation}
\label{eq:D_sp}
	D^R_{\boson}(\omega,\textbf{k})
	=
	- \frac{1}{2}[k^2 - i \alpha \, \omega + \mb^2(T)]^{-1}.
\end{equation}
Here, $\omega$ and $k$ are frequency and momentum, $\alpha$ is a constant, and $\mb^2 = \alpha_m T$ is the thermal mass. 
The latter generically arises due to symmetry-allowed quartic bosonic interactions 
\cite{
SYK_Patel_PRB_21,
SYK_Patel_linearT_arxiv_2022,
MFL_Wu_Liao_Foster_PRB_22,
thermal_mass_Torroba}. 
With disorder, the bosons behave diffusively, in contrast with the clean self-energy $\sim \omega/k$. 
Consequently, the fermionic self-energy acquires a MFL form for $T>T_*$
\cite{
MFL_Varma_CuO_PRL_89,
SYK_Patel_PRB_21,
SYK_Patel_linearT_arxiv_2022,
MFL_Wu_Liao_Foster_PRB_22}:
\begin{align}
\label{eq:MFL_self_energy}
	\Sigma^R_{\fermion}(\omega)
	=&\,
	-
	i\gammael 
	+
	\Sigma^R_{\mfl}(\omega),
\nonumber\\
	\Sigma^R_{\mfl}(\omega)
	=&\,
	-
	\gbar^2 \left(\omega \ln \frac{\omega_c}{x} + i \frac{\pi}{2} x\right),
\end{align}
in contrast with its clean counterpart that scales as $\sim |\omega|^{2/3}$. 
Here, 
$x = \max(|\omega|, J\, T)$, 
where $J$ is a slowly varying dimensionless function of $\alpha/\alpha_m$~\cite{MFL_Wu_Liao_Foster_PRB_22} 
that we take to be a constant, 
and 
$\omega_c$ is a cutoff.
The MFL self-energy indicates that the concept of quasiparticles is ill-defined due to the strong quantum-critical interaction. 
Eqs.~(\ref{eq:D_sp}) and (\ref{eq:MFL_self_energy}) are obtained in a particular large-$N$ limit \cite{SYK_Patel_linearT_arxiv_2022,MFL_Wu_Liao_Foster_PRB_22},
but the setting of fermions with strong dissipation and quantum-relaxational bosons  
in dirty strange metals is quite general.

In this paper we calculate the pairing susceptibility at finite temperature to determine the stability of the dirty MFL to superconductivity. 
Evaluating the usual Cooper ladder diagram
\cite{
NLsM6_Kamenev_CUP_11,
disO_review_PALee_RMP_1985} 
gives the strongly suppressed transition temperature in Eq.~(\ref{TcS}), due to the MFL self-energy in Eq.~(\ref{eq:MFL_self_energy}).	
This is however only the \emph{semiclassical} approximation to the pairing susceptibility. 
We also compute the Maekawa-Fukuyama diagrams 
\cite{
disO_SC_Finkelstein_Fukuyama_PhysicaB_1994,
disO_SC_Fukuyama_PhysicaB_1982} 
that describe quantum interference-mediated mixing of different interaction channels
\cite{
disO_SC_MFC_Feigelman_PRL_2007,
disO_SC_MFC_Feigelman_AnnPhy_2010,
disO_SC_MFC_Burmistrov_PRL_2012,
disO_SC_MFC_Foster_PRL_2012,
disO_SC_MFC_Burmistrov_PRB_2015}.
We will show that the quantum correction to the semiclassical result predicts robust superconductivity of
non-BCS form, Eq.~(\ref{eq:Tc}).

The model that we study here without $W$ was found to show a conductivity that diverges at low $T$~\cite{MFL_Wu_Liao_Foster_PRB_22}. 
If $W$ is generated from ultraviolet fluctuations in the boson mode, our results suggest that the low-temperature state of 
this model 
should 
be a superconductor that owes its existence to the interplay between a critical mode and disorder. 
Similar results were very recently obtained in Ref.~\cite{disO_SC_MFC_Sri_Burmistrov_arxiv_2022} 
for superconductivity near a ferromagnetic quantum critical point. 

The outline of this paper is as follows.
In Sec.~\ref{sec:results} we describe the model for a dirty superconducting marginal Fermi liquid, and we present the main results 
for the pairing susceptibility. 
In Sec.~\ref{sec:BCSMFC}, we review the multifractal enhancement of superconductivity near a metal-insulator transition 
from the perspective of the pairing of exact eigenstates. In Sec.~\ref{sec:MFCDirtMix}, we review the mixing of 
interaction channels and enhancement or suppression of pairing in dirty normal metals via the sigma model.
We precisely define the sigma model employed in this work in Sec.~\ref{sec:NLsM}. 
Details of the calculations leading to our main results appear in Sec.~\ref{sec:PairSus},
and we conclude in Sec.~\ref{sec:Conc}.

\section{Main results \label{sec:results}}

\subsection{Model}

We consider a system with $N$ flavors of fermions coupled to SU$(N)$ matrix bosons 
tuned to the quantum critical point 
\cite{
NFL_SU_N_Raghu_PRL_19,
NFL_SU_N_disO_Raghu_PRL_20,
MFL_Wu_Liao_Foster_PRB_22}. 
In the presence of disorder, the bosons become quantum relaxational and the fermions behave as in a dirty MFL 
\cite{
MFL_Wu_Liao_Foster_PRB_22,
SYK_Patel_linearT_arxiv_2022,
SYK_Guo_Patel_linearT_PRB_2022}. 
We employ the sigma-model formulation of the system
that governs its diffusive hydrodynamic modes 
\cite{
NLsM5_Finkelshtein_83,
Keldysh_conv4_Kamenev_AdvPhy_09,
NLsM2_Foster_Liao_Ann_17,
NLsM10_Burmistrov_review_19,
NLsM6_Kamenev_CUP_11}.
The theory is transcribed in the Keldysh framework
for the MFL with spin-rotational and time-reversal symmetries
(orthogonal class \cite{10-fold_Altland_Zirnbauer_PRB_1997,10-fold_Ludwig_NewJPhy_2010}).
The sigma model
can be derived by following the standard procedures of disorder averaging, 
$\hq$-matrix decoupling, and gradient expansion by assuming 
$v_F k, \omega \ll \gammael$ ($v_F$ is the Fermi velocity). 
The corresponding partition function and action are given respectively 
by 
\cite{MFL_Wu_Liao_Foster_PRB_22,NLsM2_Foster_Liao_Ann_17}
\begin{align}
	Z
	&=
	\int \calD \hq \; \calD \hphi \; \calD |\Delta|^2
	\;
	\e^{-S},
\\
	S &= S_q + S_{\phi} + S_{q\phi} + S_{\Delta} + S_{q\Delta},
\end{align}
where
\begin{align}
\label{eq:Sq_w}
S_q 
=&\,
\frac{\pi \nu_0 D}{8}
\int\limits_{\textbf{x}}
\Tr
[
	\Nabla \hq \cdot \Nabla \hq
]
\\
&\,
+
 \frac{i \pi \nu_0}{2}
\int\limits_{\textbf{x}}
\Tr
\left[
	\hq \, \hsig^3 (\homega - \hSig) 
\right],
\nonumber
\\
S_{\phi}
=&\,
-
\frac{i}{2}
\int\limits_{\omega,\textbf{k}}
\Tr
\left\{
	\hphi_{-\omega,-\textbf{k}}^{\sfT}
	[\hD_{\boson}(\omega,\textbf{k})]^{-1}
	\hphi_{\omega,\textbf{k}}
\right\},
\\
S_{q\phi}
=&\,
-
i \pi \nu_0 \, 
\frac{g}{\sqrt{N}}
\int\limits_{\textbf{x}}
\Tr
\left[
	\hat{\Phi}(\textbf{x})
	\,
	\hq(\textbf{x})
\right],
\\
S_{\Delta}
=&\,
-
i \frac{2 N}{W}
\int\limits_{t,\textbf{x}}
(
	\Deltabarq
	\Deltacl
	+
	\Deltaq 
	\Deltabarcl
),
\\
S_{q\Delta}
=&\,
\frac{i \pi \nu_0}{2}
\int\limits_{\textbf{x}}
\Tr
\left[
	\hat{\frakD}(\textbf{x})
	\,
	\hq(\textbf{x})
\right].
\end{align}
We adopted the shorthand notation $\int_{t,\textbf{x}} = \int dt \int d^2 x$ and $\int_{\omega,\textbf{k}} = \int d\omega \, d^2 k / (2\pi)^3 $.
The matrix field 
$
	\hq 
	\rightarrow 
	\hq_{\sigma,\sigma'; \alpha,\beta; i,j; \omega,\omega'}^{a,b}(\textbf{x})
$ 
incorporates both particle-hole (diffuson) and particle-particle (Cooperon) modes of the fermions,
and
carries
particle-hole $\{\sigma,\sigma'\}$,
spin-1/2 $\{\alpha,\beta\}$, 
SU($N$) flavor $\{i,j\}$,
and
Keldysh $\{a,b\}$ indices  
as well as frequency labels $\{\omega,\omega'\}$.
It obeys the constraints 
$\Tr[\hq] = 0$, 
$\hq^2 = 1$, 
and 
$\hs^2 \hsig^1 \htau^1 \hThe^1 \hq^{\sfT} \hs^2 \hsig^1 \htau^1 \hThe^1 = \hq$.
Pauli matrices  $\hsig$, $\hs$, $\htau$, and $\hThe$ act respectively in the particle-hole, spin, Keldysh, and frequency spaces \cite{NLsM2_Foster_Liao_Ann_17}.
$D = v_F^2 /4\gammael$ is the fermion diffusion constant. 
The quantum-relaxational boson is denoted $\hat{\phi}$; this is a matrix in flavor but a local field in space and time. 
Superconductivity arises via condensation of the local spin-singlet pairing fields $\{\bar{\Delta},\Delta\}$; $W$ denotes bare the BCS coupling. 

The $\hq$-matrix action differs from the Fermi-liquid case in the second line of Eq.~(\ref{eq:Sq_w}), 
where it incorporates the MFL self-energy  
$\hSig \equiv \diag\{\Sigma^R_\mfl,\Sigma^A_\mfl\}_\tau$ [Eq.~(\ref{eq:MFL_self_energy})]. 
This results in an anomalous diffusion propagator \cite{MFL_Wu_Liao_Foster_PRB_22}. 
The Yukawa and pairing interactions between $\hq$, $\hphi$, and $\Delta$  appear in  $S_{q\phi}$ and  $S_{q\Delta}$.
The fields $\hat{\Phi} \sim \hphi$ and $\hat{\frakD} \sim \Delta + \bar{\Delta}$
incorporate appropriate matrix factors and Keldysh labels $\sfcl, \sfq$, 
omitted here for brevity (see Sec.~\ref{sec:NLsM} for details).

Technically we consider flavor ``triplet'' pairing, wherein the pairing field \cite{NFL_SU_N_Wang_PRB_17}
\begin{align}\label{eq:tripletpair}
	\Delta_{\alpha,\beta; i, j} 
	=
	\Delta_0 \, \hat{s}^2_{\alpha \beta} \, \delta_{i j}.
\end{align}
SU($N > 2$) does not permit condensation of flavor-singlet fermion pairs. 
A natural generalization of our work allowing genuine flavor-singlet pairs would incorporate SO($N$) \emph{antiferromagnetic} bosons,
but this requires the further incorporation of (e.g., weakly broken) particle-hole symmetry \cite{DellAnna06,Foster08}.	

Our goal is to evaluate the retarded pairing susceptibility $\chi^R$ by integrating out the $\hq$ matrix and bosonic $\hphi$ fields
to obtain an effective propagator for the $\Delta$ field. 
We proceed by parameterizing $\hq$ with coordinates 
\cite{MFL_Wu_Liao_Foster_PRB_22,NLsM2_Foster_Liao_Ann_17} 
and derive a set of Feynman rules. 
In addition to $S_{q\phi}$ and $S_{q\Delta}$, there is a Hikami box quartic interaction term appearing after the parameterization of $S_q$. 
Diagram conventions follow Refs.~\cite{NLsM2_Foster_Liao_Ann_17,MFL_Wu_Liao_Foster_PRB_22},
see Sec.~\ref{sec:NLsM}.

\subsection{Pairing susceptibility}

\begin{figure}
\centering
\includegraphics[width=0.95\linewidth]{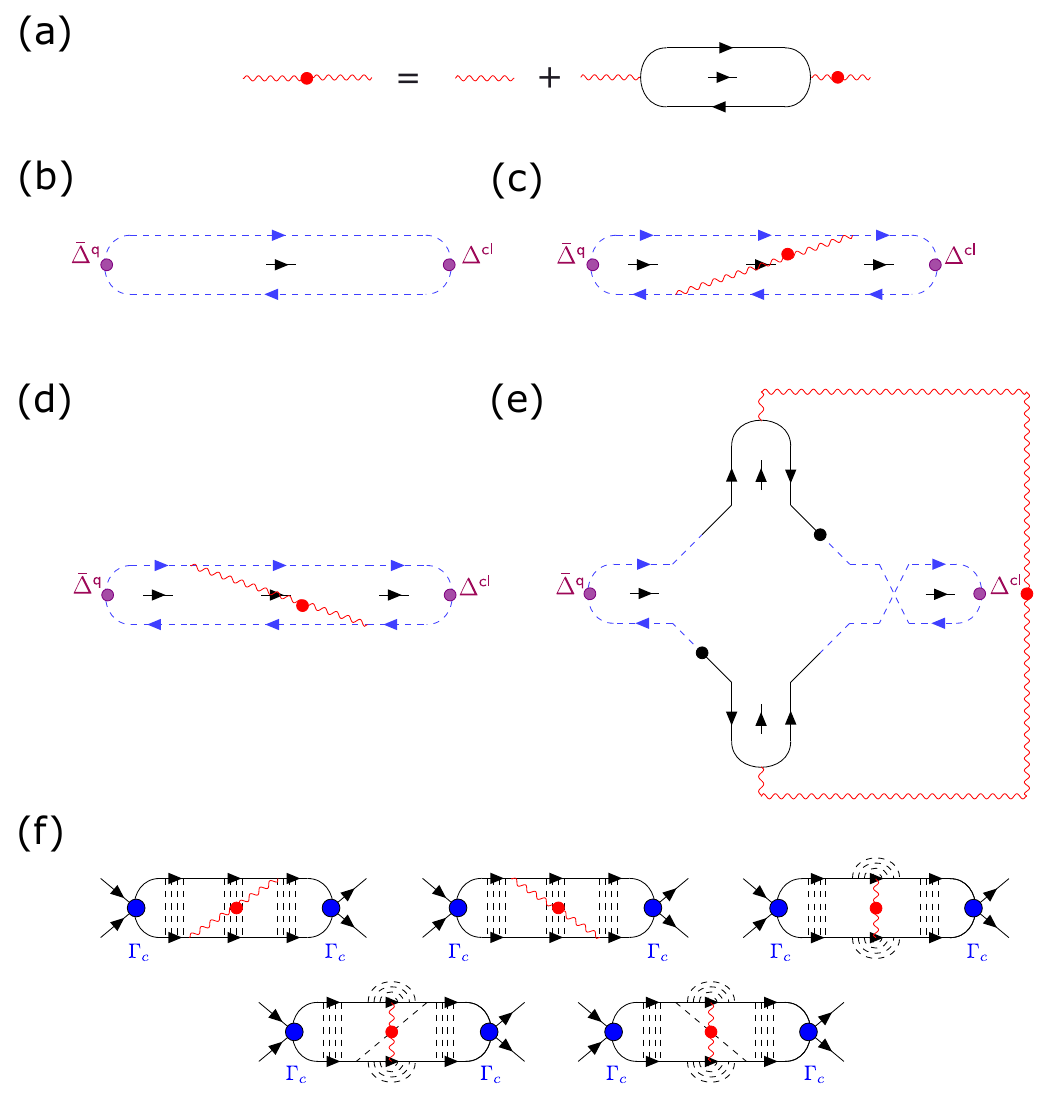}
\caption{
(a): The Feynman diagram for the dynamically screened SU($N$) matrix boson propagator (red wavy line with a red dot) under the random phase approximation. 
The wavy line represents the quantum relaxational bosonic propagator $D^R_{\boson}$ [Eq.~(\ref{eq:D_sp})]. 
(b)--(e): The diagrams contributing to the inverse static pairing susceptibility 
$[\chi^R]^{-1}(\Omega = 0,\vex{q} = \textbf{0})$: 
Panel (b) is the semiclassical Cooperon ladder contribution, see also Eq. (\ref{eq:cal_S}).
Panels (c)--(e) are the quantum interference corrections due to the interplay of diffusive collective modes (diffusons and Cooperons) 
and the attractive quantum-critical interaction, see Eq.~(\ref{eq:chi_total}).
The diffuson propagator is represented diagrammatically by two black
solid lines carrying counterpropagating arrows.
Similarly, the Cooperon propagator is represented by two blue dashed lines. 
$\Delta^{\sfcl/\sfq}$ represents the Cooper pairing field. 
The vertex for the coupling between $\Delta^{\sfcl/\sfq}$ and the Cooperon is represented by the purple dot. 
More details appear in Secs.~\ref{sec:NLsM} and \ref{sec:PairSus}.   
(f): The Feynman diagrams renormalizing the BCS channel scattering amplitude $\Gamma_c$ in conventional many-body perturbation theory \cite{disO_SC_Finkelstein_Fukuyama_PhysicaB_1994,disO_SC_Fukuyama_PhysicaB_1982}. 
These quantum corrections are analogous to diagrams (c)--(e) within the sigma model. 
The dashed lines represent impurity scattering and solid black lines denote the fermionic Green's functions. 
}
\label{fig:pairing_susceptibility}
\end{figure}

\begin{figure}
\centering
\includegraphics[width=0.9\linewidth]{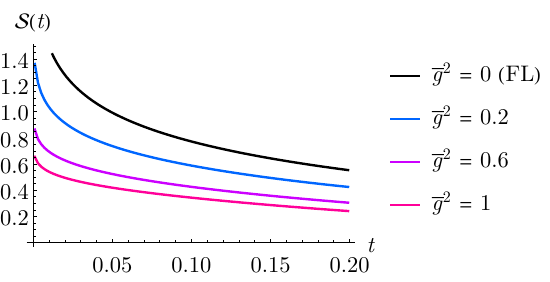}
\caption{
Plot of the semiclassical Cooperon ladder integral $\calS(T)$ in Eq.~(\ref{eq:cal_S}) as a function of the dimensionless temperature $t = T/\gammael$, 
for different reduced fermion-boson Yukawa couplings $\gbar^2$ (increasing from black to red). 
Here, the dephasing rate due to the Cooperon self-energy is chosen to be $\tau_{C}^{-1}/\gammael = 10^{-3}$.  
The black curve corresponds to the Fermi liquid case in which there is no marginal Fermi liquid (MFL) self-energy contribution. 
The singularity of $\calS$ is strongly suppressed by the MFL effects as $\gbar^2$ increases,
owing to the absence of well-defined quasiparticles.
}
\label{fig:calS}
\end{figure}

At the semiclassical level, the Cooperon ladder contribution to the pairing susceptibility 
\cite{
NLsM6_Kamenev_CUP_11,
disO_review_PALee_RMP_1985} 
is diagrammatically depicted in Fig.~\ref{fig:pairing_susceptibility}(b). 
The corresponding (retarded) inverse pairing susceptibility as $T \rightarrow T_c$ is
\begin{equation}
	[\chi^R_{\sfsemi}]^{-1}(\Omega \rightarrow 0, \textbf{q} \rightarrow 0)
	=
	-
	\frac{2 N}{W}
	+
	2 \pi \nu_0 \,N\, \calS(t_c)
	=
	0,
\end{equation}
where
\begin{align}
\label{eq:cal_S}
	\calS(t)
	&=
	\int_{-1/t}^{1/t}
	\frac{dy}{2\pi}
	\frac{
		\tanh (y)
	}{
		 y  \,
		\calA_{\sfMFL}(y) 
		+
		i/(4t \gammael\, \tau_{\vphi})
	}
	,
\\
	\calA_{\sfMFL}(y) 
	&= 
	1 + \gbar^2 \ln\left[ \frac{\omega_c / T}{\max(J,2|y|)} \right],
\end{align}
$t \equiv T / \gammael$ 
and 
$\tau_{\vphi}^{-1} \equiv \tau_{\sfMFL}^{-1} + \tau_{C}^{-1}$ is the dephasing rate that in principle contains effects from both the MFL and the Cooperon self-energy \cite{NLsM2_Foster_Liao_Ann_17,dephasing_Liao_Foster}. 
To make the physical picture more transparent, we will treat $\tau_{C}^{-1}$ as a phenomenological constant in this work for simplicity and defer a more detailed analysis to a separate study.  
The temperature dependence of $\calS$ for various $\gbar^2$ is shown in Fig.~\ref{fig:calS}. 
At low temperatures, the dephasing rate drops out and $\calS$ can be estimated as
\begin{equation}
	\calS(t)
	\simeq
	\frac{1}{\pi \, \gbar^2}
	\ln (1 - \gbar^2 \ln t_c).
\end{equation}
In the absence of the MFL self-energy ($\gbar^2 = 0$, black curve), $\calS$ diverges logarithmically as $T$ decreases, resulting in the standard BCS result 
$T_c^{\mathsf{BCS}} \propto \e^{-1/\nu_0 W}$ for $1/T_c \tau_{\vphi} \ll 1$, independent of disorder and in accordance with Anderson's theorem. 
However, as $\gbar^2$ increases (blue to red), the logarithmic divergence is increasingly weakened by $\calA_{\sfMFL} > 1$, leading to a much smaller 
transition temperature
$T_c^{\mathcal{S}}$ in Eq.~(\ref{TcS}).
The suppression can be understood physically as a consequence of attempting to pair incoherent MFL fermions \cite{NFL_Raghu_BCS_PRB_15}. 
Although the MFL self-energy does \emph{not} contribute to transport at the semiclassical level with a spatially uniform Yukawa coupling $g$ \cite{MFL_Wu_Liao_Foster_PRB_22,SYK_Patel_linearT_arxiv_2022,SYK_Guo_Patel_linearT_PRB_2022}, 
it strongly degrades superconductivity as there is no conservation law associated with the Cooperon propagator.

We now consider quantum interference corrections to the pairing susceptibility at the leading order of $1/\sigma_{\sfdc}$. 
The corresponding Feynman diagrams are shown in Figs.~\ref{fig:pairing_susceptibility}(c)--(e). 
These involve the dynamically screened quantum-critical boson, which has the propagator
\begin{align}\label{eq:Dscr}
	D^R_{\sfscr}(\omega,\textbf{k})
	&=
	\frac{1}{
		[D^R_{\boson}]^{-1}(\omega,\textbf{k})
		-
		\Pi^R_{\boson}(\omega,\textbf{k})
	},
	\\
	\Pi^R_{\boson}(\omega,\textbf{k})
	&\simeq
	\frac{2i}{N}
	\frac{
		\nu_0 g^2  \omega
	}{
		Dk^2 - i \omega \, \left[1 + \gbar^2 \ln(\omega_c/|\omega|) \right]
	},
\end{align}
where $D^R_{\boson}$ is given by Eq. (\ref{eq:D_sp}). 
Depending on the temperature $T$, the quantum correction scales differently due to the competition between thermal and dynamical screening. 
The explicit expressions for the diagram amplitudes can be found in Sec.~\ref{sec:PairSus}.
The quantum diagrams in Figs.~\ref{fig:pairing_susceptibility}(c)--(e) are analogous to those considered by  
Maekawa and Fukuyama \cite{disO_SC_Fukuyama_PhysicaB_1982,disO_SC_Finkelstein_Fukuyama_PhysicaB_1994}. 
For readers unfamiliar with this notation, the analogous diagrams in conventional many-body perturbation theory are shown in Fig.~\ref{fig:pairing_susceptibility}(f). 
For the case of repulsive Coulomb interactions, we recover the $\ln^3(\Lambda/T_c)$ suppression of $T_c$ \cite{disO_SC_Finkelstein_Fukuyama_PhysicaB_1994,disO_SC_Feigelman_PRB_2000,disO_SC_Fukuyama_PhysicaB_1982}.
  
Upon summing up the diagrams, the total inverse pairing susceptibility in the static limit is found to be 
\begin{align}
\label{eq:chi_total}
	[\chi^R_{\mathsf{tot}}]^{-1}(0,\textbf{0})
	&=
	-
	\frac{2 N}{W}
	+
	\pi \nu_0 \,  
	\left[
		N \calS(t)
		+
		{\cal Q}(t)
	\right].
\end{align}
The quantum correction
\begin{equation}
	{\cal Q}(t)
	\simeq
	\begin{cases}
		\dfrac{  \gbar^2 \, \calC(t)}{\pi \nu_0 \, D\, t}, \quad 
		& t \gtrsim t_*,
		\\
		7.25   \, \sqrt{\dfrac{N\, \gbar^2}{2^3 \pi^4 \nu_0 \, D \, t}}, 	
		& t \ll  t_*,
	\end{cases}
\end{equation}
where $t = T/\gammael$ is the reduced temperature and $t_* \equiv 2\pi^2 \gbar^2 \nu_0/ (\alpha_m N)$. 
For $ t\gtrsim t_*$, thermal screening dominates and the singularity in $t$ originates from the thermal mass of the quantum-relaxational bosons [Eq.~(\ref{eq:D_sp})]. 
For $t \ll t^*$, the thermal mass and dephasing drop out and the quantum correction diverges in a less singular manner. 
$\calC(t)$ is a dimensionless parameter which varies weakly with $T$ and depends on $\alpha$, $\alpha_m$, $\gbar^2$ and $\tau_{\vphi}^{-1}$ (see Sec.~\ref{sec:PairSus}). 
It encodes the integrals involved in Figs.~\ref{fig:pairing_susceptibility}(c)--(e). 

The second and third terms in Eq.~(\ref{eq:chi_total}) correspond respectively to the semiclassical and quantum corrections. 
As the $T$-dependence of the semiclassical contribution $\calS(T)$ is extremely weak, its primary effect is merely to enhance the BCS coupling $W$ to $W_{\sfeff}$:
\begin{align}\label{eq:Weff}
	W_{\sfeff}^{-1} \equiv W^{-1} - (\pi \nu_0/2) \, \calS(t_c). 
\end{align} 
Meanwhile, the quantum correction is singular in $t$ owing to the quantum-critical nature of the bosons. 
Consequently, superconductivity is mainly driven by the quantum correction. 
As $T \rightarrow T_c$, $[\chi^R_{\mathsf{tot}}]^{-1}(0,\textbf{0}) \rightarrow 0$. Solving for $T_c$  using Eq.~(\ref{eq:chi_total}) results in Eq.~(\ref{eq:Tc}).
The enhancement due to the quantum correction overcomes the MFL suppression and increases as $\gbar^2$.
This is the main result of this paper.

\subsection{Discussion}

The physical origin of the pairing enhancement is two-fold.
On one hand, the quantum-critical interaction is magnified by the collective modes encoding the diffusive motion of the fermions. 
Physically, this is because the fermions have a higher probability to interact strongly with each other due to their slow diffusive motion \cite{NLsM6_Kamenev_CUP_11,disO_review_PALee_RMP_1985,disO_review_Kirkpatrick_94}. 
This in turns amplifies the interaction in the BCS channel through operator mixing that results in multifractal enhancement 
\cite{disO_SC_MFC_Burmistrov_PRL_2012,disO_SC_MFC_Burmistrov_PRB_2015}. 
On the other hand, as the attractive SU($N$) Altshuler-Aronov conductance correction suppresses localization, fermions become more itinerant at low $T$ \cite{MFL_Wu_Liao_Foster_PRB_22,NFL_SU_N_disO_Raghu_PRL_20}. 
These two effects overcome the MFL suppression of the single-particle phase coherence and promote superconductivity through collective excitations. 
Our results differ from the suppressive Maekawa-Fukuyama quantum correction 
\cite{disO_SC_Fukuyama_PhysicaB_1982,disO_SC_Finkelstein_Fukuyama_PhysicaB_1994,disO_SC_Feigelman_PRB_2000}, 
owing to the attractive and long-ranged nature of the quantum-critical interaction studied here. 

Our work suggests a novel mechanism to boost superconductivity in a disordered quantum-critical system. 
Importantly, the robust superconductivity predicted by Eq.~(\ref{eq:Tc}) 
demonstrates how collective phase coherence can survive in systems without well-defined quasiparticles. 
Similar conclusions apply to Altshuler-Aronov quantum corrections to the conductivity \cite{Maslov05,Galitski05,Ludwig08,NFL_SU_N_disO_Raghu_PRL_20,MFL_Wu_Liao_Foster_PRB_22}. 
While the above calculation was carried out perturbatively in $1/\sigma_{\sfdc}$, the physical picture is more general 
and we expect it to hold when higher-order corrections are included.

\section{REVIEW: Multifractal enhancement of superconductivity \label{sec:BCSMFC}}

In this section, we review the key steps leading to the prediction of enhanced superconductivity in an ordinary (not strange) metal due to static quantum-critical spatial fluctuations near an Anderson metal-insulator transition (MIT). Following Refs.~\cite{disO_SC_MFC_Feigelman_PRL_2007,disO_SC_MFC_Feigelman_AnnPhy_2010}, we consider the pairing of exact eigenstates of a disordered, noninteracting system. 
Although this simple approach demonstrates that wave-function multifractality can enhance pairing, it is less accurate than a full solution to the self-consistent Bogoliubov-de Gennes equations 
\cite{disO_SC_Num_Trivedi_PRL_1998,disO_SC_Num_Trivedi_PRB_2001}. Enhanced superconductivity with full self-consistency has been demonstrated in 
Refs.~\cite{disO_SC_MFC_Garcia-Garcia_PRB_2020,disO_SC_MFC_Zhang_Foster_PRB_2022}.

We consider an attractive Hubbard model with disorder in $d$ spatial dimensions,
\begin{align}\label{Hubbard}
	H
	=
	-
	\sum_{i,j}	
	t_{i,j}
	\,
	c^\dagger_{i \sigma}
	c_{j \sigma}
	-
	|U|
	\sum_{i}
	c_{i \uparrow}^\dagger c_{i \downarrow}^\dagger c_{i \downarrow} c_{i \uparrow}.
\end{align}
Here $c_{j \sigma}$ annihilates an electron at site $j$ with spin $\sigma$, and 
$t_{i,j}$ is the single-particle Hamiltonian that encodes hopping and quenched disorder. 
We assume time-reversal symmetry, so that exact eigenstates of $\hat{t}$ arise in degenerate pairs.   
Let $\psi_\alpha(i)$ denote such an eigenstate with eigenenergy $\xi_\alpha$.
We re-express Eq.~(\ref{Hubbard}) in this basis,  
\begin{alignat}{2}\label{Hubbard-PoEE}
	H
	=&\,
	\sum_{\alpha}	
	\xi_\alpha
	\,
	c_{\alpha \sigma}^\dagger c_{\alpha \sigma}
\nonumber\\
&\,
	-
	|U|
	\sum_{\alpha,\beta,\gamma,\nu}
&&\,
\left[
	\sum_{i}
	\psi_{\alpha}^*(i)
	\psi_{\beta}^*(i)
	\psi_\gamma(i)
	\psi_\nu(i)
\right]
\nonumber\\
&
&&\,
	\times
	c_{\alpha \uparrow}^\dagger c_{\beta \downarrow}^\dagger c_{\gamma \downarrow} c_{\nu \uparrow}.
\end{alignat}
As in the reduced BCS approximation, we simplify the interaction term, retaining only products of Cooper-pair annihilation
and creation operators that populate or empty a pair of single-particle states.
The mates of the pair are related by time-reversal. 
In the absence of spin-orbit coupling, 
we get
\begin{align}
	H_{\mathsf{red}}
	=
	\sum_{\alpha}	
	\xi_\alpha
	\,
	c_{\alpha \sigma}^\dagger c_{\alpha \sigma}
	-
	\sum_{\alpha,\beta}
	M_{\alpha,\beta}
	\,
	c_{\alpha \uparrow}^\dagger c_{\bar{\alpha} \downarrow}^\dagger c_{\bar{\beta} \downarrow} c_{\beta \uparrow},
\end{align}
where the labels $(\bar{\beta},\downarrow)$ denote the time-reversed partner of $(\beta,\uparrow)$.
The eigenstate correlations are encoded in the matrix element
\begin{align}
	M_{\alpha \beta}
	=
	|U|
	\sum_{i}
	\left|\psi_\alpha(i)\right|^2
	\left|\psi_\beta(i)\right|^2.
\end{align} 
Defining a BCS order parameter $\Delta_\alpha \equiv - \sum_{\beta} M_{\alpha \beta} \, c_{\bar{\beta}\downarrow} c_{\beta \uparrow}$ and following
the standard steps, we obtain the BCS mean-field equation 
\begin{align}\label{BCS}
	\Delta_\alpha
	=
	\sum_{\beta}
	M_{\alpha \beta}
	\tanh\left(\frac{E_\beta}{2 T}\right)
	\frac{\Delta_\beta}{2 E_\beta},
\end{align}
where $T$ denotes the temperature and $E_{\beta} = \sqrt{\xi_\beta^2 + |\Delta_\beta|^2}$.

For weak pairing near the Fermi energy in a system residing within the diffusive metallic phase or close to the MIT (but still on the metallic side),
we can take $\Delta_\alpha \rightarrow \Delta$, independent of $\alpha$. Then Eq.~(\ref{BCS}) can be rewritten in terms of an energy integration, 
\begin{align}\label{BCSMFC}
	\frac{1}{|U|}
	\simeq
	\left(a^d \nu_0\right)^{1 - \Upsilon}
	\int
	\frac{d \xi}{|\xi|^{\Upsilon}}
	\frac{
	\tanh\left(
		\frac{\sqrt{\xi^2 + \Delta^2}}{2 T}
	\right)
	}{2 \sqrt{\xi^2 + \Delta^2}}.
\end{align}
Here $\nu_0$ is the density of states and $a$ is the lattice spacing. 
The exponent $\Upsilon \equiv (d - d_2)/d$ measures the \emph{second multifractal dimension} $d_2$ of the eigenstates near the Fermi energy.
In the metallic phase $d_2 = d$ and Eq.~(\ref{BCSMFC}) reduces to the usual BCS form. 
Near the MIT, however, multifractal eigenstates have $0 < d_2 < d$, leading to an infrared enhancement of the integral. 
Neglecting any UV cutoff (the integral is now convergent), Eq.~(\ref{BCSMFC}) yields an amplified, non-BCS
prediction for $T_c \sim |U|^{1/\Upsilon}$. A more realistic treatment 
\cite{disO_SC_MFC_Garcia-Garcia_PRB_2015}
retains the Debye energy cutoff to the 
integral in Eq.~(\ref{BCSMFC}).

\section{REVIEW: Disorder-mediated mixing of interaction channels \label{sec:MFCDirtMix}}

The approach in Sec.~\ref{sec:BCSMFC} cannot be easily generalized to the case of a strange metal, where quasiparticles are ill-defined due to Planckian dissipation. 
In this paper, we instead employ the sigma model, which allows the incorporation of non-Fermi liquid effects \cite{NFL_SU_N_disO_Raghu_PRL_20,MFL_Wu_Liao_Foster_PRB_22}.
In this section, we review the sigma-model version of disorder-induced interaction operator mixing in a diffusive Fermi liquid
\cite{disO_SC_MFC_Burmistrov_PRL_2012,disO_SC_MFC_Burmistrov_PRB_2015}. Results are presented for two spatial dimensions,
obtained by calculating Feynman diagrams involving interaction operators and interference processes. The latter are treated to lowest nontrivial order in $1/\sigma_{\sfdc}$, where $\sigma_{\sfdc}$ is the dimensionless semiclassical conductivity. By taking appropriate limits, we can demonstrate (a) the enhancement of superconductivity, consistent with the results of the previous section, or (b) the suppression of $T_c$, due to long-ranged Coulomb interactions \cite{disO_SC_Fukuyama_PhysicaB_1982,disO_SC_Finkelstein_Fukuyama_PhysicaB_1994}.

In the sigma-model approach to a diffusive Fermi liquid, renormalization group (RG) beta functions can be derived for dimensionless interaction couplings. 
For a time-reversal invariant system with strong spin-orbit coupling, the interactions include density-density and BCS pairing coupling strengths $\gamma_s$ and $\gamma_c$, respectively.
Repulsive (attractive) interactions in each channel correspond to $\gamma > 0$ ($\gamma < 0$). 
The lowest-order beta functions are 
\begin{align}\label{floweq}
	\frac{d }{d l}
	\begin{bmatrix}
	\gamma_s
	\\
	\gamma_c
	\end{bmatrix}
	=
	-
	\frac{\lambda}{2}
	\begin{bmatrix}
		1 & -2 \\
		-1 & 0 
	\end{bmatrix}
	\begin{bmatrix}
	\gamma_s
	\\
	\gamma_c
	\end{bmatrix}
	-
	\begin{bmatrix}
	0
	\\
	2 \gamma_c^2
	\end{bmatrix},
\end{align}
where $l = \ln(L)$ is the logarithm of the RG lengthscale (e.g.\ system size $L$) and $\lambda \propto 1/\sigma_{\sfdc}$.
The second term on the right-hand side (RHS) of Eq.~(\ref{floweq}) drives the usual BCS instability, which survives in the presence of non-magnetic disorder (Anderson's theorem).
The first term on the RHS of Eq.~(\ref{floweq})  encodes the renormalization of the interactions due to disorder. The matrix is off-diagonal because the bare density-density and pairing terms are not eigenoperators of the interference processes. 

Ignoring the nonlinear $-2 \gamma_c^2$ term for the moment, the flow equations in (\ref{floweq}) possess a single relevant direction that drives both interactions towards strong coupling. 
For suitable initial conditions, this results in an instability wherein both $\gamma_{s,c} \rightarrow - \infty$. 
We denote $\alpha \equiv - 4 \gamma_{s,c}/3$ (tuning the irrelevant difference of the interaction coupling strengths to zero), and we then obtain a beta function for the instability
\begin{align}
	\frac{d \alpha}{d l} = \eta \, \alpha + \alpha^2,
\end{align}
where $\eta = \lambda/2$. We solve this equation to determine $\alpha = \alpha(L)$. At a certain length scale $L$ the coupling diverges; we identify 
this scale with the coherence length $\zeta_\mathsf{coh}$ of the superconducting state, 
leading to a ground-state gap of the form
\begin{align}
	\Delta
	\sim
	\alpha_0^{1 / \Upsilon},
\end{align}
where $\alpha_0$ is the bare coupling and $\Upsilon = \eta / d$. Here we have used the fact that $\Delta \sim \zeta_\mathsf{coh}^{-d}$ for superconductivity nucleating from a diffusive Fermi liquid in $d$ spatial dimensions. 
At an Anderson MIT, $\Upsilon$ becomes a universal scaling exponent, leading to a prediction consistent the results presented in the last section. 

The flow equations (\ref{floweq}) apply for a Fermi liquid with short-ranged density-density and BCS-pairing interactions. For a charged liquid with long-ranged Coulomb interactions, one should 
instead pin $\gamma_s \rightarrow 1$ \cite{disO_SC_MFC_Burmistrov_PRB_2015}. In this case, the remaining flow equation for the Cooper coupling reduces to
\begin{align}
	\frac{d \gamma_c}{d l}
	=
	\eta
	-
	2 \gamma_c^2.
\end{align}
The first term on the RHS of this equation corresponds to a perturbative shift of $\gamma_c$ to the repulsive side,
\begin{align}
	\delta \gamma_c = \eta \ln\left(\frac{L}{a}\right),
\end{align}
which obtains from the diagrams in 
Figs.~\ref{fig:pairing_susceptibility}(c)--(f).
For an overall attractive $\gamma_c < 0$, the change in $T_c$ is 
\begin{align}\label{MF}
	\frac{\delta T_c}{T_c^{\mathsf{BCS}}}
	\sim&\,
	-
	\frac{\delta \gamma_c}{\gamma_c^2}
	=
	-
	\eta 
	\ln\left(\frac{L}{a}\right)
	\,
	\ln^2\left(\frac{\Lambda}{T_c^{\mathsf{BCS}}}\right)
\nonumber\\
	\sim&\,
	-
	\frac{\eta}{d} 
	\,
	\ln^3\left(\frac{\Lambda}{T_c^{\mathsf{BCS}}}\right),
\end{align}
where again we have converted $T \sim L^{-d}$.
Eq.~(\ref{MF}) is the Maekawa-Fukuyama suppression of $T_c$ due to interference in a charged, diffusive Fermi liquid  \cite{disO_SC_Fukuyama_PhysicaB_1982,disO_SC_Finkelstein_Fukuyama_PhysicaB_1994}.

\section{The marginal-Fermi-liquid non-linear sigma model \label{sec:NLsM}}

We consider $s$-wave superconductivity 
with spin-singlet, flavor ``triplet'' pairing [Eq.~(\ref{eq:tripletpair})],
in a system of disordered fermions with $N$ flavors coupled to SU($N$) quantum-critical bosons 
\cite{
NFL_SU_N_Raghu_PRL_19,
NFL_SU_N_disO_Raghu_PRL_20,
MFL_Wu_Liao_Foster_PRB_22}.   
In the presence of spin rotational and time-reversal symmetries, the system in the diffusive regime 
can be described by the class-AI Finkelstein non-linear sigma model (FNLsM), which is an effective field 
theory describing the diffusive collective modes of the fermions
\cite{
Keldysh_conv4_Kamenev_AdvPhy_09,
NLsM2_Foster_Liao_Ann_17,
MFL_Wu_Liao_Foster_PRB_22,
NLsM5_Finkelshtein_83,
disO_review_Kirkpatrick_94,
NLsM10_Burmistrov_review_19}. 
The derivation of the sigma model is similar to the one in class A. 
Upon disorder averaging and integrating out the fermions, followed by the gradient expansion 
we obtain the following partition function and action 
\cite{
NLsM2_Foster_Liao_Ann_17,
MFL_Wu_Liao_Foster_PRB_22}, 
\begin{widetext}
\begin{align}
	Z
	&=
	\int \calD \hq \; \calD \hphi \; \calD |\Delta|^2
	\;
	\e^{-S}
\\
\label{eq:action_S}
	S
	&=
	\frac{1}{8\lambda}
	\int\limits_{\textbf{x}}
	\Tr
	[
		\Nabla \hq \cdot \Nabla \hq
	]
	+
	i \frac{h}{2}
	\int\limits_{\textbf{x}}
	\Tr
	\left\{
		\hq \left[\hsig^3 (\homega - \hSig) + i\eta \, \hsig^3 \htau^3\right]
	\right\}
\\
	&
	-
	\frac{i}{2}
	\int\limits_{\Omega,\textbf{k}}
	\Tr
	\left\{
		\hphi_{-\Omega,-\textbf{k}}^{\sfT}
		[\hD_{\boson}(\Omega,\textbf{k})]^{-1}
		\hphi_{\Omega,\textbf{k}}
	\right\}
	-
	ih \frac{g}{\sqrt{N}}
	\int\limits_{\textbf{x}}
	\Tr
	\left[
		\left(
			\hphi^a \htau^a \hEsig_{11} + \hs^2 \hphi^{a \sfT}\htau^a \hs^2 \hEsig_{22}
		\right)
		\hM_{F}(\homega)
		\hq(\textbf{x})
		\hM_{F}(\homega)
	\right]
	\nonumber
\\
	&+
	\frac{ih}{2}
	\int\limits_{\textbf{x}}
	\Tr
	\left[
			\left(
				\Deltacl \hsig^+
				+
				\Deltabarcl \hsig^-
				+
				\Deltaq \hsig^+ \htau^1
				+
				\Deltabarq \hsig^- \htau^1
			\right)
		\hM_F(\homega)
		\hq(\textbf{x})
		\hM_F(\homega)
	\right]
	-
	i \frac{2 N}{W}
	\int\limits_{t,\textbf{x}}
	(
		\Deltabarq
		\Deltacl
		+
		\Deltaq 
		\Deltabarcl
	)
	.
	\nonumber
\end{align}
\end{widetext}
The Pauli matrices $\htau$, $\hsig$, $\hat{s}$, and $\hThe$ act respectively in the Keldysh, particle-hole, spin-1/2, and frequency spaces. 
$\hThe^1$ is defined via $\bra{\omega} \hThe^1 \ket{\omega'} = 2\pi \,  \delta(\omega + \omega')$, where $\omega$ is frequency. 
The thermal matrix $\hM_{F}$ is defined as
\begin{equation}
	\hM_{F}(\homega) 
	=
	\begin{bmatrix}
		1 & F(\homega)
		\\
		0 & - 1
	\end{bmatrix}_{\tau},
\end{equation}
where $F(\omega) = \tanh(\omega/2T)$ is the generalized Fermi distribution function and $T$ denotes temperature. 
The subscript $\tau$ means that this is a matrix in the Keldysh space. 
For brevity, we work with the units $k_B = c = \hbar = 1$, where $k_B$ is the Boltzmann constant, $c$ is the speed of light, and $\hbar$ is the Planck constant. 
The $\hq$ matrix 
$
	\hq 
	\rightarrow 
	\hq_{\sigma,\sigma';i,j;\omega_1,\omega_2;\alpha,\beta}^{a,b}(\textbf{x})
$ 
carries indices in
$\{\sigma,\sigma'\} \in \{1,2\}$ particle-hole,
$\{\alpha,\beta\} \in \left\lbrace \uparrow,\downarrow \right\rbrace$ spin-1/2,
$\{i,j\} \in \left\lbrace 1,2,...,N \right\rbrace$ flavor,
and 
$\{a,b\} \in \{R,A\}$ Keldysh (fermion retarded/advanced \cite{NLsM2_Foster_Liao_Ann_17}) spaces,
as well as frequency labels $\{\omega_1,\omega_2\}$. 
It is subjected to the following constraints
\begin{equation}
	\hq^2 = 1,
	\qquad
	\Tr[\hq] = 0,
	\qquad
	\hs^2 \hsig^1 \htau^1 \hThe^1 \hq^{\sfT} \hs^2 \hsig^1 \htau^1 \hThe^1 = \hq.
\end{equation}

In Eq.~(\ref{eq:action_S}),
we have employed the shorthand notations 
$\int_{t,\textbf{x}} = \int dt \int d^2 x$ 
and 
$\int_{\Omega,\textbf{q}} = \int \frac{d\Omega}{2\pi} \int \frac{d^2 q}{(2\pi)^2}$. 
We have also introduced the particle-hole-space projectors 
	$\hEsig_{11} 
	\equiv
	(1 + \hsig^3)/2
	$
	,
	$\hEsig_{22} 
	\equiv
	(1 - \hsig^3)/2
	$, 
and
	$\hsig^{\pm} \equiv (\hsig^1 \pm i\, \hsig^2)/2$. 
The coupling constants for the pure $\hat{q}$-matrix action are
\begin{equation}
\label{eq:coupling_const}
	h = \pi \nu_0
	,
	\qquad
	\frac{1}{\lambda} = D h
	,
\end{equation}
where $\nu_0$ is the density of states per channel,
the diffusion constant is $D = v_F^2/(4\gammael)$, 
and $\gammael$ is the elastic impurity scattering rate.
The small dimensionless parameter $\pi \lambda$ is the inverse conductance per channel.

In the above action $S$, we have Hubbard-Stratonovich (H.S.) decoupled the bare pairing interaction 
in the spin-singlet, 
flavor-``triplet'' [Eq.~(\ref{eq:tripletpair})],
$s$-wave channel, resulting in the H.S.\ fields $\Delta^{\sfcl,\sfq}$ and $\bar{\Delta}^{\sfcl,\sfq}$. 
$W$ in the last term of Eq.~(\ref{eq:action_S}) is the effective Bardeen-Cooper-Schrieffer (BCS) interaction strength. 

The last term on the second line of Eq.~(\ref{eq:action_S}) describes the linear coupling between the 
fermion-bilinear $\hq$ matrix and the matrix bosonic field $\hphi \rightarrow \phi^a_{ij}(\Omega,\textbf{k})$, 
which carries Keldysh label $a = \sfcl, \sfq$ and flavor indices $\{i,j\}$. 
This term mediates the $S_{\sfintI}$ interaction in Eq.~(77) of Ref.~\cite{MFL_Wu_Liao_Foster_PRB_22}.
The Keldysh matrices $\{\htau^{\sfcl},\htau^{\sfq}\} \equiv \{\hat{1},\htau^1\}$. 
 
The SU($N$) boson propagator $\hD_{\boson}$ at the saddle-point level \cite{MFL_Wu_Liao_Foster_PRB_22} 
is given by 
\begin{equation}
\begin{aligned}
	\hD_{\boson}(\Omega,\textbf{k}) 
	=&\,
	\begin{bmatrix}
	D^K_{\boson}(\Omega,\textbf{k}) & D^R_{\boson}(\Omega,\textbf{k})
	\\
	D^A_{\boson}(\Omega,\textbf{k}) & 0
	\end{bmatrix}
	,
\\
	D^R_{\boson}(\Omega,\textbf{k})
	=&\,
	-
	\frac{1}{2
	(
		\textbf{k}^2 - i \alpha \, \Omega + \alpha_m T
	)
	}
	,
\end{aligned}
\end{equation}
where $\alpha$ and $\alpha_m$ are constants. 
The $\alpha_m T$ term serves as the thermal mass for the quantum-critical bosons and generically arises due to symmetry-allowed quartic $\phi^4$ interactions amongst the bosons 
\cite{
MFL_Wu_Liao_Foster_PRB_22,
SYK_Patel_linearT_arxiv_2022,
SYK_Guo_Patel_linearT_PRB_2022,
thermal_mass_Torroba}. 
In contrast with the $|\omega|/k$ frequency structure in the clean case, the bosonic propagator now behaves 
diffusively due to disorder smearing. Such quantum-relaxational form of the bosonic propagator is quite generic 
and was also obtained in a slightly different model 
\cite{
SYK_Guo_Patel_linearT_PRB_2022,
SYK_Patel_linearT_arxiv_2022}. 

The quantum-relaxational boson induces a marginal Fermi liquid (MFL) self-energy for the fermions
\cite{MFL_Varma_CuO_PRL_89,SYK_Guo_Patel_linearT_PRB_2022,SYK_Patel_linearT_arxiv_2022,MFL_Wu_Liao_Foster_PRB_22}
(instead of $\sim |\omega|^{2/3}$), 
which enters the sigma model as the matrix  
\begin{subequations}\label{MFL}
\begin{equation}
	\hSig_{\homega}
	=
	\begin{bmatrix}
	\hSig_{\mfl,\homega}^R & 0 
	\\
	0 & \hSig_{\mfl,\homega}^A
	\end{bmatrix}_{\tau}
	=
	\re \hSig_{\mfl,\homega}^R
	+
	i
	\,
	\im \hSig_{\mfl,\homega}^R 
	\,
	\htau^3,
\end{equation}
where
\begin{equation}
	\Sigma_{\mfl,\omega}^R =
	-\gbar^2 \left[\omega \ln\left(\frac{\omega_c}{x}\right) + i \frac{\pi}{2} x\right]
	,
	\quad
	x = \max(|\omega|, J T).
\end{equation}
\end{subequations}
Here $J$ is a constant and $\gbar^2 = g^2/4\pi^2 \gammael$ is the dimensionless squared Yukawa coupling constant. 
Note that the causality structure of the $\hsig^3 \im \hSig_{\mfl, \homega}$ term in the sigma model is consistent with 
the $i\eta \, \hsig^3 \tau^3$ Keldysh prescription. 

The $S_{\sfintII}$ term in Eq.~(78) of Ref.~\cite{MFL_Wu_Liao_Foster_PRB_22} is a vertex-correction term that is crucial
for the calculation of the density response function, in order to satisfy the Ward identity for charge conservation. 
It also prevents strange metallicity (linear-$T$ resistivity) from manifesting 
\emph{at the semiclassical level} 
for
pure potential disorder \cite{SYK_Patel_linearT_arxiv_2022,MFL_Wu_Liao_Foster_PRB_22}.
However
$S_{\sfintII}$ is irrelevant for the discussion of the pairing susceptibility and we thus do not discuss this term further in this paper. 

In the following, we present the parameterization of the FNLsM and the corresponding Feynman rules. Our main goal is to compute the pairing susceptibility and explore the consequences of the interplay between disorder and quantum-critical interactions on the superconducting transition temperature $T_c$. 
While we focus on a particular microscopic model here, the physics of the conclusion is quite general and expected to be applicable to a wide range of disordered quantum-critical systems.

\subsection{Rotation of the saddle point and  the $\pi-\sigma$ parameterization}

To facilitate the parameterization, we perform a unitary rotation \cite{NLsM2_Foster_Liao_Ann_17}
\begin{equation}
	\hq \rightarrow \hR \, \hq  \, \hR,
	\qquad
	\hR
	=
	\hEsig_{11}
	+
	\hEsig_{22} \, \htau^1
	,
\end{equation} 
where $\hEsig_{11,22}$ are defined above Eq.~(\ref{eq:coupling_const}), 
so that the saddle point becomes
\begin{equation}
	\hqsp \rightarrow \htau_3.
\end{equation}
The symmetry constraints for $\hq$ are now
\begin{equation}
\label{eq:q_constrain}
	\hq^2 = 1, 
	\qquad 
	\Tr[\hq] = 0, 
	\qquad 	
	\hs^2 \hsig^1 \hThe^1 \hq^{\sfT}  \hs^2 \hsig^1 \hThe^1 = \hq.
\end{equation}
We then employ the ``$\pi$-$\sigma$'' parameterization for the $\hq$ matrix
\begin{align}
\label{eq:q_pi_sigma_para}
	\hq
	=&\,
	\begin{bmatrix}
		\sqrt{
			1 - \hWdag \hW
		}
		& \hWdag
		\\
		\hW
		&
		-\sqrt{
			1 - \hW \hWdag
		}
	\end{bmatrix}_{\tau}
\nonumber\\
	=&\,
	\hq^{(0)}
	+
	\hq^{(1)}
	+
	\hq^{(2)}
	+
	\hq^{(4)}
	+
	\ldots
\end{align}
where 
\begin{eqnarray}
	\hq^{(0)} &=& \htau^3,
	\\
	\hq^{(1)} 
	&=&
	\begin{bmatrix}
		0 & \hWdag
		\\
		\hW & 0
	\end{bmatrix}_{\tau},
	\\
	\hq^{(2)}
	&=&
	\frac{1}{2}
	\begin{bmatrix}
		-\hWdag \hW & 0
		\\
		0 & \hW \hWdag
	\end{bmatrix}_{\tau},
	\\
	\hq^{(4)}
	&=&
	\frac{1}{8}
	\begin{bmatrix}
		-(\hWdag \hW)^2 & 0 
		\\
		0 & (\hW \hWdag)^2
	\end{bmatrix}_{\tau}
	,
\end{eqnarray}
and the $\hW$ field satisfies
\begin{equation}
\label{eq:constraint_W}
	\hW = \hs^2 \hsig^1 \hThe^1 (\hWdag)^{\sfT} \hs^2 \hsig^1 \hThe^1,
\end{equation}
which follows immediately from Eq. (\ref{eq:q_constrain}).

We further introduce the unconstrained,
$\mathbb{C}$-valued
matrix fields $\hX$ and $\hY$ defined by
\begin{align}
\begin{aligned}
	X_{ij;\omega_1,\omega_2;\alpha \beta}(\textbf{k})
	=&\,
	W^{+,+}_{ij;\omega_1,\omega_2;\alpha\beta}(\textbf{k}),
\\
	Y_{ij;\omega_1,\omega_2;\alpha \beta}(\textbf{k})
	=&\,
	W^{+,-}_{ij;\omega_1,\omega_2;\alpha \beta}(\textbf{k}).
\end{aligned}
\end{align} 
To satisfy the constraint in Eq. (\ref{eq:constraint_W}), we can write
\begin{equation}
\label{eq:def_W}
	\hW 
	=
	\begin{bmatrix}
		\hX & \hY 
		\\
		\hThe^1 \hs^2 (\hYdag)^{\sfT}\hThe^1\hs^2  & \hThe^1 \hs^2 (\hXdag)^{\sfT} \hThe^1 \hs^2
	\end{bmatrix}_{\sigma}
	.
\end{equation}
We then plug Eqs.~(\ref{eq:q_pi_sigma_para}) and (\ref{eq:def_W}) back into the FNLsM and 
expand the action in Eq.~(\ref{eq:action_S}) up to quartic order.  

To facilitate the perturbative expansion in $\lambda$, which is proportional to inverse conductance, we rescale
$\hX \rightarrow \sqrt{\lambda} \, \hX$,
$	\hY \rightarrow \sqrt{\lambda} \, \hY$
\cite{NLsM2_Foster_Liao_Ann_17,MFL_Wu_Liao_Foster_PRB_22}.
At quadratic order, the action is given by
\begin{widetext}
\begin{eqnarray}
	S_X^{(2)}
	&=&
	\frac{1}{2}
	\int
	\Tr
	\left[
		\hXdag_{1,2}(\textbf{k}_1)
		M_{2,1;4,3}(\textbf{k}_1,\textbf{k}_2)
		\hX_{3,4}(\textbf{k}_2)
		+
		\Idag_{2,1}(\textbf{k})
		\hX_{1,2}(\textbf{k})
		+
		I_{2,1}(\textbf{k})
		\hXdag_{1,2}(\textbf{k})
	\right]
	,
	\\
	S_Y^{(2)}
	&=&
	\frac{1}{2}
	\int
	\Tr
	\left[
		\hYdag_{1,2}(\textbf{k}_1)
		N_{2,1;4,3}(\textbf{k}_1,\textbf{k}_2)
		\hY_{3,4}(\textbf{k}_2)
		+
		\Jdag_{2,1}(\textbf{k})
		\hY_{1,2}(\textbf{k})
		+
		J_{2,1}(\textbf{k})
		\hYdag_{1,2}(\textbf{k})
	\right]
	,
\end{eqnarray}
where the numerical subscripts $\left\lbrace 1,2,3,4\right\rbrace$ correspond to frequencies $\left\lbrace \omega_1,\omega_2,\omega_3,\omega_4\right\rbrace $, $\int = \int_{1,2,3,4,\textbf{k}}$ integrates over all frequencies and momentum, and 
$\Tr$ traces over the flavor and spin-1/2 labels. 
We also defined 
\begin{equation}
	\begin{aligned}
		M_{2,1;4,3}(\textbf{k}_1,\textbf{k}_2)
		=
		\left[\Delta_{2,1}^X(\vex{k}_1)\right]^{-1}
		\,
		\delta_{1,4}\delta_{2,3} \delta_{\textbf{k}_1,\textbf{k}_2}
		&+
		ih\lambda  \frac{1}{\sqrt{N}}
		[ 
			\hphicl_{4-1}(\textbf{k}_1 - \textbf{k}_2)
			+
			F_4 \hphiq_{4-1}(\textbf{k}_1 - \textbf{k}_2)
		]
		\,
		\delta_{2,3}
		\\
		&+
		ih\lambda  \frac{1}{\sqrt{N}}
		[
			-\hphicl_{2-3}(\textbf{k}_1 - \textbf{k}_2)
			+
			F_3 \hphiq_{2-3}(\textbf{k}_1 - \textbf{k}_2)
		]
		\,
		\delta_{1,4}
		,
	\end{aligned}
\end{equation}
\begin{equation}
	\begin{aligned}
		N_{2,1;4,3}(\textbf{k}_1,\textbf{k}_2)
		=
		\left[\Delta_{2,1}^Y(\vex{k}_1)\right]^{-1}
		\,
		\delta_{1,4}\delta_{2,3} \delta_{\textbf{k}_1,\textbf{k}_2}
		&+
		ih\lambda \frac{1}{\sqrt{N}}
		[
			\hphicl_{4-1}(\textbf{k}_1 - \textbf{k}_2)
			-
			F_1 \hphiq_{4-1}(\textbf{k}_1 - \textbf{k}_2)
		]
		\,
		\delta_{2,3}
		\\
		&+
		ih\lambda  \frac{1}{\sqrt{N}}
		[
			-\hphicl_{2-3}(\textbf{k}_1 - \textbf{k}_2)
			+
			F_3 \hphiq_{2-3}(\textbf{k}_1 - \textbf{k}_2)
		]
		\,
		\delta_{1,4}
		,
	\end{aligned}
\end{equation}
\begin{eqnarray}
	\Idag_{2,1}(\textbf{k})
	&=&
	2i h \sqrt{\lambda}  \frac{1}{\sqrt{N}}
	[
		(F_2 - F_1) \hphicl_{2-1}(-\textbf{k})
		+
		(1 - F_1 F_2) \hphiq_{2-1}(-\textbf{k})
	]
	,
	\\
	I_{2,1}(\textbf{k})
	&=&
	2i h \sqrt{\lambda}   \frac{1}{\sqrt{N}}\hphiq_{2-1}(\textbf{k})
	,
	\\
	\Jdag_{2,1}(\textbf{k})
	&=&
	2 i h \sqrt{\lambda}  \frac{1}{\sqrt{N}}
	\left(
		\Deltabarcl_{1-2} (\textbf{k})
		-
		F_1 \Deltabarq_{1-2} (\textbf{k})
	\right)
	,
	\\
	J_{2,1}(\textbf{k})
	&=&
	2i h \sqrt{\lambda}  \frac{1}{\sqrt{N}}
	\left(
		\Deltacl_{2-1}(\textbf{k}) - F_1 \Deltaq_{2-1} (\textbf{k})
	\right)
	,
\end{eqnarray}
where the Dirac delta functions $\delta_{\omega_1,\omega_2} \equiv \delta_{1,2} \equiv 2\pi \, \delta(\omega_1 - \omega_2)$ and $\delta_{\textbf{k}_1,\textbf{k}_2} = (2\pi)^2 \, \delta^{(2)}(\textbf{k}_1 - \textbf{k}_2)$. 

The bare propagators are given by
\begin{eqnarray}
\label{XXProp}
	\left\langle 
		X_{i,j; 1,2; \alpha, \beta}(\textbf{k})
		\,
		\Xdag_{k,l;  3,4 ;\delta, \gamma}(\textbf{k})
	\right\rangle_0
	&=&
	2\Delta_{1,2}^X(\textbf{k})\, \delta_{1,4} \, \delta_{2,3} \, \delta_{il} \, \delta_{jk}\, \delta_{\alpha \gamma} \delta_{\beta,\delta},
\\
\label{YYProp}
	\left\langle 
		Y_{i.j;1,2;\alpha, \beta}(\textbf{k})
		\,
		\Ydag_{k,l;3,4;\delta, \gamma}(\textbf{k})
	\right\rangle_0
	&=&
	2\Delta_{1,2}^Y(\textbf{k}) \, \delta_{1,4} \, \delta_{2,3}\, \delta_{il} \, \delta_{jk}\, \delta_{\alpha \gamma} \delta_{\beta,\delta},
\\
	\Delta_{1,2}^X(\textbf{k})
	&=&
	\frac{1}{
		k^2 + ih \lambda (\omega_1 - \omega_2) + ih \lambda \gbar^2(\omega_1 \ln \frac{\omega_c}{x_1} - \omega_2 \ln \frac{\omega_c}{x_2} ) + h \lambda \, \tau_{\mathsf{MFL}}^{-1}
	}
	,
\\
	\Delta_{1,2}^Y(\textbf{k})
	&=&
	\frac{1}{
		k^2 + ih \lambda (\omega_1 + \omega_2) + ih \lambda \gbar^2(\omega_1 \ln \frac{\omega_c}{x_1} + \omega_2 \ln \frac{\omega_c}{x_2} ) + h \lambda \,  \tau_{\mathsf{MFL}}^{-1}
	}
	,
\end{eqnarray}
where $x_{1,2} = \max(|\omega_{1,2}|,J T)$ 
and 
$\tau_{\mathsf{MFL}}^{-1} = \gbar^2 \pi J \, T$ is the dephasing rate due to the MFL self-energy $\hSig_{\mfl}$
[Eq.~(\ref{MFL})].
Note: the prefactor 2 in Eqs.~(\ref{XXProp}) and (\ref{YYProp})
is due to the prefactor $1/2$ in the Gaussian action.

At quartic order, there are five Hikami box terms \cite{NLsM2_Foster_Liao_Ann_17}.
The relevant one responsible for the pairing susceptibility to the leading order of $\lambda$ is
\begin{equation}
	S_q^{(4)}
	=
	\frac{\lambda}{4}
	\int
	\delta_{\textbf{k}_1 + \textbf{k}_3,\textbf{k}_2 +\textbf{ k}_4}
	\Box^{\textbf{k}_1,\textbf{k}_2,\textbf{k}_3,\textbf{k}_4}_{1,2,3,4}
	\Tr
	\begin{bmatrix}
		\hX_{1,2}(\textbf{k}_1)
		\hs^2 \hY^{\sfT}_{3,-2}(-\textbf{k}_2)
		\hX^{\dagger \sfT}_{4,3}(-\textbf{k}_3) \hs^2
		\Ydag_{-4,1}(\textbf{k}_4)
	\end{bmatrix}
	,
\end{equation}
where $\Tr$ traces over the flavor and spin indices, 
$\sfT$ is the transpose in the flavor $\otimes$ spin space, 
and
\begin{equation}
	\Box_{1,2,3,4}^{\textbf{k}_1,\textbf{k}_2,\textbf{k}_3,\textbf{k}_4}
	=
	\left[
	\begin{aligned}
	&\,
		-
		(\textbf{k}_1 \cdot \textbf{k}_3 + \textbf{k}_2 \cdot \textbf{k}_4)
		+
		\frac{1}{2}
		(\textbf{k}_1 + \textbf{k}_3)
		\cdot
		(\textbf{k}_2 + \textbf{k}_4)
	\\&\,
		+
		i
		\frac{h}{2}
		\lambda\left(
			\omega_1 - \omega_2 + \omega_3 - \omega_4 
			- \Sigma_{\mfl,1}^R 
			+ \Sigma_{\mfl,2}^A 
			- \Sigma_{\mfl,3}^R 
			+ \Sigma_{\mfl,4}^A
		\right)
	\end{aligned}
	\right].
\end{equation}

\begin{figure}
\centering
\includegraphics[width=0.35\linewidth]{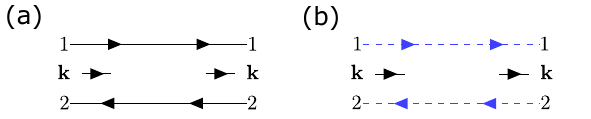}
\caption{
The diagrammatic representation of the propagators for 
(a) diffusons ($\hX$ fields) 
and 
(b) Cooperons ($\hY$ fields). 
The arrows on these lines represent frequency flow in the propagators. 
The numerical labels $1,2$ are shorthand notations for frequencies $\omega_1$, $\omega_2$.  
Meanwhile, the short arrows between the two lines indicate the momentum (\textbf{k}) flow. 
}
\label{fig:propagator}
\end{figure}

\subsection{Feynman rules}

We adopt the same convention for the Feynman rules presented in Refs.~\cite{NLsM2_Foster_Liao_Ann_17,MFL_Wu_Liao_Foster_PRB_22}. 
The diffuson (Cooperon) propagator is represented diagrammatically in Figs.~\ref{fig:propagator} (a) [(b)] by two black
solid lines (blue dashed lines) with arrows pointing in the opposite directions. 
The frequency indices of the matrix fields $\hX$ ($\hY$) and $\hXdag$ ($\hYdag$) are labeled by the numbers. 
The flavor and spin indices are implicit. 
Along a solid line, the frequency label, spin, and flavor indices remain unchanged. 
The momentum flowing through the diffuson is labeled by an arrow in the middle of the two black solid lines. 
The momentum arrow points inwards (outwards) for the fields $\hX, \hY$ ($\hXdag, \hYdag$). 

The interaction vertices are diagrammatically depicted in Fig.~\ref{fig:vertex_all}. 
The corresponding amplitudes are given by

\begin{figure}
\centering
\includegraphics[width=0.7\linewidth]{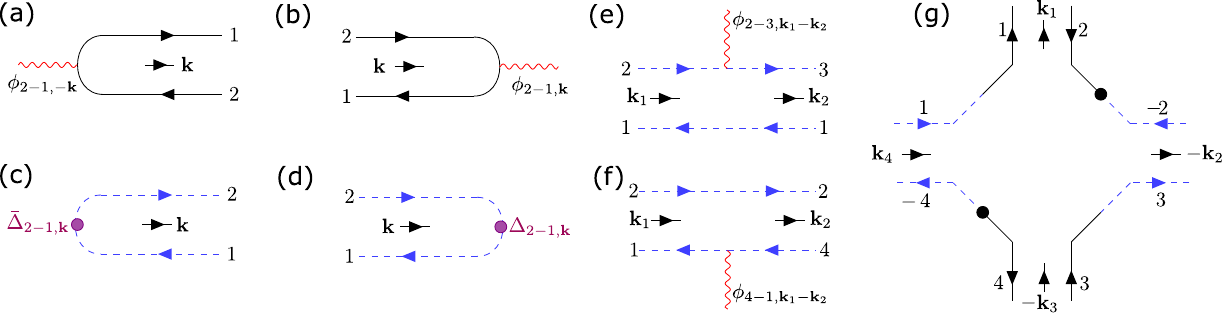}
\caption{
The Feynman diagrams representing the vertices relevant for evaluating the pairing susceptibility 
[see Eqs.~(\ref{eq:feynman_rule_a})--(\ref{eq:feynman_rule_f})]. 
The diffuson (Cooperon) propagator is represented diagrammatically by two black
solid (blue dashed) lines with arrows pointing in the opposite directions (see Fig.~\ref{fig:propagator}). 
The red wavy line denotes the bosonic field $\phi$.  
The vertex for converting between the Hubbard-Stratonovich Cooper pairing field $\Delta^{\sfcl/\sfq}$ 
and the Cooperon mode is denoted by the purple dot. 
}
\label{fig:vertex_all}
\end{figure}

\begin{eqnarray}
\label{eq:feynman_rule_a}
	\text{Fig. }\ref{fig:vertex_all} \text{(a)}
	&=&
	-ih\sqrt{\lambda}
	\,
	[
		(F_2 - F_1)\phicl_{2-1,-\textbf{k}}
		+
		(1 - F_1 F_2)\phiq_{2-1,-\textbf{k}}
	]
	/
	\sqrt{N}
	,
	\\ 
\label{eq:feynman_rule_b}
	\text{Fig. }\ref{fig:vertex_all} \text{(b)}
	&=&
	-ih\sqrt{\lambda}
	\phiq_{2-1,\textbf{k}}
	/
	\sqrt{N}
	,
	\\
\label{eq:feynman_rule_c}
	\text{Fig. } \ref{fig:vertex_all} \text{(c)}
	&=& 
	-i h \sqrt{\lambda}
	\,
	[
		\Deltabarcl_{1-2,\textbf{k}} 
		-
		F_1 \Deltabarq_{1-2,\textbf{k}} 
	],
	\\
\label{eq:feynman_rule_d}
	\text{Fig. } \ref{fig:vertex_all} \text{(d)} 
	&=& 
	- i h \sqrt{\lambda}
	\,
	[
		\Deltacl_{2-1,\textbf{k}} - F_1 \Deltaq_{2-1,\textbf{k}}
	],
	\\
\label{eq:feynman_rule_e}
	\text{Fig. }\ref{fig:vertex_all} \text{(e)}
	&=& 
	- \frac{ih}{2}\lambda
	\,
	[
		-\phicl_{2-3,\textbf{k}_1- \textbf{k}_2}
		+
		F_3 \phiq_{2-3,\textbf{k}_1- \textbf{k}_2}
	]
	/
	\sqrt{N}
	,
	\\
\label{eq:feynman_rule_f}
	\text{Fig. }\ref{fig:vertex_all} \text{(f)}
	&=& 
	-\frac{ih}{2}
	\lambda
	\,
	[
		\phicl_{4-1,\textbf{k}_1- \textbf{k}_2}
		-
		F_1 \phiq_{4-1,\textbf{k}_1- \textbf{k}_2}
	]
	/
	\sqrt{N}
	,
	\\
	\text{Fig. }\ref{fig:vertex_all} \text{(g)}
	&=&
	-
	\frac{\lambda}{4}
	\Box_{1,2,3,4}^{\textbf{k}_1,\textbf{k}_2,\textbf{k}_3,\textbf{k}_4}
	.
\end{eqnarray}

\section{Pairing susceptibility calculation \label{sec:PairSus}}

\subsection{Semiclassical contribution - The Cooper ladder}

The retarded inverse pairing susceptibility $[\chi^R_{\sfsemi}]^{-1}$ can be obtained by computing the effective propagator of 
$\Deltacl$ and $\Deltabarq$. 
At the semiclassical level (ignoring quantum interference), we only consider the linear coupling between $\hY, \hYdag$ and $\Delta, \bar{\Delta}$. 
By integrating out the $\hY$ fields, we have 
\begin{equation}
\label{eq:chi_semi}
	i[\chi^R_{\sfsemi}]^{-1}(\omega,\textbf{k})
	=
	-
	i
	\frac{2N}{W}
	+
	4 h^2 \lambda N
	\int\limits_{\veps}
	\frac{
		F_{\veps - \frac{\omega}{2}}
	}{
		k^2 - 2ih\lambda \veps + ih\lambda \gbar^2\left[(-\veps + \omega/2) \ln \frac{\omega_c}{|-\veps + \omega/2|} - (\veps + \omega/2) \ln \frac{\omega_c}{|\veps + \omega/2|} \right] + h\lambda \tau_{\vphi}^{-1}
	},
\end{equation}
where $\tau_{\vphi}^{-1}$ is the effective dephasing rate incorporating $\tau_{\mathsf{MFL}}^{-1}$ and the Cooperon self-energies represented by the diagrams in Fig.~18 of Ref.~\cite{NLsM2_Foster_Liao_Ann_17}. 
The corresponding Feynman diagram for the second term in Eq.~(\ref{eq:chi_semi}) is depicted in \text{Fig.~\ref{SM--fig:pairing_susceptibility}(a)}, which is equivalent to summing the Cooper ladders \cite{AA4_Altshuler_Review_85}. 
In the static limit $\omega = 0$ and $k \rightarrow 0$, we have
\begin{equation}
	i
	[\chi^R_{\sfsemi}]^{-1}(0,\textbf{0})
	=
	-
	i
	\frac{2 N}{W}
	+
	4
	h  N
	\int\limits_{\veps}
	\frac{
		F_{\veps}
	}{
		  -2i\veps  \AMFL(\veps) + \tau_{\vphi}^{-1}
	}
	,
\end{equation}
where the marginal Fermi liquid (MFL) self-energy [Eq.~(\ref{MFL})] is encoded by the factor
\begin{equation}
	\AMFL(\veps) =  1 + \gbar^2 \ln \frac{\omega_c}{\max(JT,|\veps|)} > 1.
\end{equation}
In the absence of the MFL effects, $\gbar^2 \rightarrow 0$, and the above expression reduces to the FL result \cite{Keldysh_conv4_Kamenev_AdvPhy_09,NLsM6_Kamenev_CUP_11}. 
At the proximity of the transition temperature $T \rightarrow T_c$, $[\chi^R_{\sfsemi}]^{-1} \rightarrow 0$. 
If we assume $\tau_{\vphi}^{-1} \ll T_c$ for simplicity, then
\begin{align}
0
=
[\chi^R_{\sfsemi}]^{-1}(0,\textbf{0})
&=
-
\frac{2 N}{W}
+
\frac{2 h N}{\pi}\int_0^{\Lambda/2T_c} dy
\frac{
	\tanh (y)
}{
 y  \AMFL(2T_c y)
}
\\
&\simeq
-
\frac{2 N}{W}
+
\frac{2 h N}{\pi}
\frac{1}{\gbar^2}
\ln
\left(
1 + \gbar^2 \ln \frac{\Lambda}{T_c}
\right)
,
\end{align}
implying that [Eq.~(\ref{TcS})]
\begin{equation}
	T_c
	\simeq
	\Lambda
	\,
	\exp
	\left[
		- \frac{1}{\gbar^2}\left( e^{\pi  \gbar^2/hW} - 1\right)
	\right], 
\end{equation}
which is much smaller than the BCS result. 
Physically, the strong suppression of the transition temperature 
is due to the MFL self-energy that destroys well-defined quasiparticles before they can form Cooper pairs.

\begin{figure}
\centering
\includegraphics[width=0.8\linewidth]{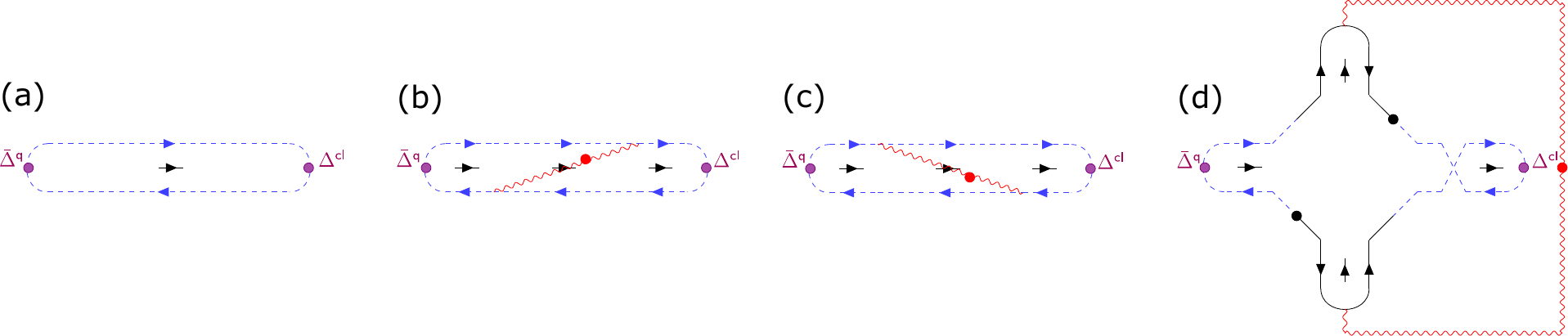}
\caption{
The Feynman diagrams responsible for the pairing susceptibility.
(a): the semiclassical contribution, 
and 
(b)--(d): 
the quantum interference corrections to the leading order in  $\lambda$, 
which is proportional to the inverse conductance.
}
\label{SM--fig:pairing_susceptibility}
\end{figure}

\subsection{Quantum interference correction}

Despite the absence of well-defined quasiparticles, phase coherence can still manifest through collective modes. 
In the following, we consider the quantum correction to the pairing susceptibility represented by the diagrams in \text{Figs.~\ref{SM--fig:pairing_susceptibility}(b)--(d)}. 
The corresponding expressions are respectively given by
\begin{align}
	\text{Fig. \ref{SM--fig:pairing_susceptibility}(b)}
	&=
	4 h^4 \lambda^3 g^2 
	\int\limits_{\veps_1,\veps_2,\textbf{q}}
	\Delta_{\veps_1,\veps_1}^Y(\textbf{k})
	\Delta_{\veps_2,\veps_2}^Y(\textbf{k})
	\Delta_{\veps_1, \veps_2}^Y(\textbf{k} + \textbf{q})
	(-F_{\veps_1^+})
	i
	\left[
		F_{\veps_2^+} D^A_{\sfb,\veps_1 - \veps_2}(\textbf{q})
		+
		F_{\veps_1^-} D^R_{\sfb,\veps_1 - \veps_2}(\textbf{q})
	\right],
\label{Fukuyama1}
\\
	\text{Fig. \ref{SM--fig:pairing_susceptibility}(c)}
	&=
	4 h^4 \lambda^3 g^2  
	\int\limits_{\veps_1,\veps_2,\textbf{q}}
	\Delta_{\veps_1,\veps_1}^Y(\textbf{k})
	\Delta_{\veps_2,\veps_2}^Y(\textbf{k})
	\Delta_{\veps_1, \veps_2}^Y(\textbf{k} + \textbf{q})
	(-F_{\veps_1^+})
	i
	\left[
		F_{\veps_1^-} D^R_{\sfb,\veps_1 - \veps_2}(\textbf{q})
		+
		F_{\veps_2^+} D^A_{\sfb,\veps_1 - \veps_2}(\textbf{q})
	\right],
\label{Fukuyama2}
\\
	\text{Fig. \ref{SM--fig:pairing_susceptibility}(d)}
	&=
	8 h^4 \lambda^3 
	g^2
	\int\limits_{\veps_1,\veps_2,\textbf{q}}
	\Delta_{\veps_1,\veps_1}^Y(\textbf{k})
	\Delta_{\veps_2,\veps_2}^Y(\textbf{k})
	[\Delta_{\veps_1, -\veps_2}^X(\textbf{q})]^2
	\Box_{\veps_1^+,-\veps_2^+, \veps_2^-, -\veps_1^-}^{\textbf{q},-\textbf{k},-\textbf{q},\textbf{k}}
	F_{\veps_1^+}(-F_{\veps_1} - F_{\veps_2})
	\,
	iD^R_{\sfb,\veps_1 + \veps_2}(\textbf{q}),
\label{Fukuyama3}
\end{align}
where $\veps^{\pm} = \veps \pm \omega/2$. 
Due to the Cooperon self-energy, $\tau_{\mathsf{MFL}}^{-1}$ in the propagator $\Delta^Y$ should be replaced by 
the total dephasing rate
$\tau_{\vphi}^{-1} = \tau_{\sfMFL}^{-1} + \tau_C^{-1}$. 
In the above expressions, we omitted terms proportional to the Keldysh component of bosonic propagator; 
these can be absorbed into the dephasing rate \cite{NLsM2_Foster_Liao_Ann_17}.

\subsubsection{Thermal-screening dominated regime}

For $T \gtrsim T_*$, thermal screening is more important than dynamical screening and thus the bosonic propagator can be approximated by the quantum relaxational version 
$D_{\boson}$ [Eq.~(\ref{eq:D_sp})]. 
Here the thermal-to-dynamical screening crossover temperature is defined via
\begin{align}
	T^* \equiv \frac{g^2 \nu_0}{2 \alpha_m N},
\end{align}
where $\alpha_m = \mb^2 / T$ is the coefficient of the quantum-relaxational boson thermal mass. 

In the thermal screening regime, by focusing on the static limit $\omega \rightarrow 0$, $k \rightarrow 0$ and performing the $q$ integral, we have
\begin{equation}
	\text{Fig. \ref{SM--fig:pairing_susceptibility}(b)} + \text{Fig. \ref{SM--fig:pairing_susceptibility}(c)}  + \text{Fig. \ref{SM--fig:pairing_susceptibility}(d)}
	\simeq
	- i  \, h\lambda  \, \calC \, \frac{\gbar^2}{t},
\end{equation}
where  $t = T/\gammael$, 
and
the dimensionless parameter 
\begin{equation}
\label{eq:cal_C}
\begin{aligned}
	\calC(t)
	=&\,
	-
	\frac{ h }{\pi}
	\int
	dy_1 \, dy_2
	\,
	\frac{
		\tanh y_1
	}{
		[2i y_1 \AMFL(2T y_1) + \frac{1}{2T \tau_{\vphi}}]
		[2i y_2 \AMFL(2T y_2) + \frac{1}{2T \tau_{\vphi}}]
	}
	\\
	&
	\phantom{-}\qquad
	\times
	\begin{bmatrix}
		\Phi(y_1,y_2,\alpha,\alpha_m,\tau_{\vphi}^{-1},\lambda)
		\tanh y_1
		\\\\
		+
		\Phi(y_1,y_2,-\alpha,\alpha_m,\tau_{\vphi}^{-1},\lambda)
		\tanh y_2
		\\\\
		+
		\Phi_H(y_1,y_2,\alpha,\alpha_m,\tau_{\mathsf{MFL}}^{-1},\lambda) (\tanh y_1 + \tanh y_2)
	\end{bmatrix}
	,
\end{aligned}
\end{equation}
with
\begin{equation}
\begin{aligned}
	\Phi(y_1,y_2,\alpha,\alpha_m,\tau_{\vphi}^{-1},\lambda,\gbar^2,T)
	\equiv&\,
	-2 T
	\int_0^{\infty} \, dx
	\,
	\Delta_{2T y_1,2T y_2}^Y(x)
	D^R_{\sfb,2T(y_1 - y_2)}(x)
	\\
	=&\,
	\frac{1}{
		-ih\lambda[2 y_1 \AMFL(2T y_1)+ 2 y_2 \AMFL(2T y_2) ] - h\lambda \frac{\tau_{\vphi}^{-1}}{T} +  \alpha_m   - 2i\alpha  (y_1 - y_2)
	}
	\\
	&
	\,
	\times
	\ln
	\left\{ 
	\frac{
		\alpha_m   - 2i\alpha  (y_1  - y_2)
	}{
		ih\lambda[2 y_1 \AMFL(2T y_1)+ 2 y_2 \AMFL(2T y_2)] + h\lambda \frac{\tau_{\vphi}^{-1}}{T}
	}
	\right\}
	,
\end{aligned}
\end{equation}
and similarly
\begin{equation}
\begin{aligned}
	\Phi_H(y_1,y_2,\alpha,\alpha_m,\tau_{\sfMFL}^{-1},\lambda,\gbar^2,T)
	\equiv&\,
	-2 T
	\int_0^{\infty}  \, dx
	\,
	\Delta_{2T y_1,-2T y_2}^X(x)
	D^R_{\sfb,2T(y_1 + y_2)}(x)
	\\
	=&\,
	\frac{1}{
		-ih\lambda[2 y_1 \AMFL(2T y_1)+ 2 y_2 \AMFL(2T y_2) ] - h\lambda \frac{\tau_{\sfMFL}^{-1}}{T} +  \alpha_m   - 2i\alpha  (y_1 + y_2)
	}
	\\
	&
	\,	
	\times
	\ln
	\left\{ 
	\frac{
		\alpha_m   - 2i\alpha  (y_1 + y_2)
	}{
		ih\lambda[2 y_1 \AMFL(2T y_1)+ 2 y_2 \AMFL(2T y_2)] + h\lambda \frac{\tau_{\sfMFL}^{-1}}{T}
	}
	\right\}.
\end{aligned}
\end{equation}
Note that since the MFL and total dephasing rates, $\tau_{\sfMFL}^{-1}$ and $\tau_{\vphi}^{-1}$ respectively, are proportional to temperature, 
$\calC$ is just a function of $\ln T$ and thus only weakly depends on temperature. 
We plot $\calC(t)$ versus the dimensionless temperature $t = T/\gammael$ in Fig.~\ref{fig:plotc} for various model parameters.

\begin{figure}
\centering
\includegraphics[width=1\linewidth]{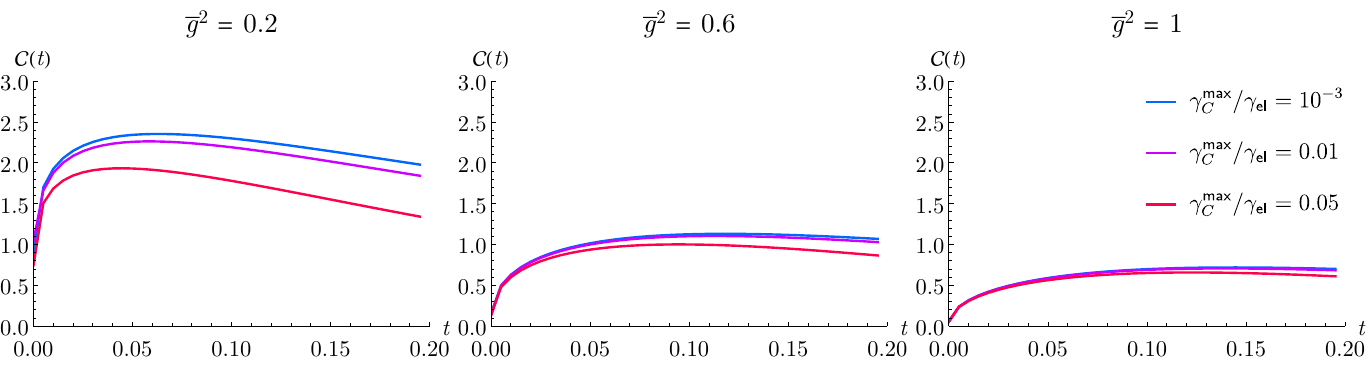}
\caption{
Plot of $\calC(t)$ as a function of the reduced temperature $t = T/\gammael$ in the thermal-screening dominated regime, based on Eq.~(\ref{eq:cal_C}). Here, the parameters are $h = 1$, $\lambda = 0.02$, $\alpha_m = \alpha = 0.5$, $J = 1$, $\Lambda = \gammael = 10$ and $\gamma_C = \tau_C^{-1}$ is the dephasing rate. 
We phenomenologically take a linear-$T$ dephasing rate $\gamma_C(t) = \gamma_C^{\sfmax} (t/t_{\sfmax})$, where $t_{\sfmax} = 0.2$.  
}
\label{fig:plotc}
\end{figure}

\subsubsection{Dynamical-screening dominated regime}

We now consider the regime $T \ll T_*$ in which dynamical screening dominates over thermal screening. 
Repeating the above calculation using the dynamically screened bosonic propagator $D_{\sfscr}$ [Eq.~(\ref{eq:Dscr})], we obtain a new version of $\Phi$. 
The full expression is very complicated, but for small $T$, 
\begin{equation}
\Phi_{\sfscr}(y_1, y_2)
=
\frac{\pi \sqrt{T}}{
2
\sqrt{2i \, (y_2 - y_1) \, \beta}
}
+
{\cal O}(T)
\end{equation}
where $ \beta = 2h^2 g^2 \lambda / \pi N$. 
This contribution arises from the regime where $q^2 \sim 2Ty_{1,2}$,
where $q$ is the loop momentum in Eqs.~(\ref{Fukuyama1})--(\ref{Fukuyama3}).
The dimensionless parameter $\calC$ then becomes
\begin{equation}
\calC_{\sfscr}
\simeq
-
\frac{h}{\pi}
\int dy_1 \, dy_2
\,
\frac{\tanh y_1}{
(2i y_1)(2i y_2)
}
\begin{bmatrix}
( \Phi_{\sfscr}(y_1, y_2) \tanh y_1 +  \Phi_{\sfscr}^*(y_1, y_2) \tanh y_2)
\\\\
+\Phi_{\sfscr}(y_1, -y_2) (\tanh y_1 + \tanh y_2).
\end{bmatrix}
\end{equation}
By numerically evaluating the remaining integral, we have
\begin{equation}
\re
\calC_{\sfscr}
\simeq
7.25 
\frac{h}{\pi}
\sqrt{\frac{T}{\beta}}
=
7.25 
\frac{h}{\pi}
\sqrt{\frac{T}{2h^2 g^2 \lambda / \pi N}}
=
7.25 
\sqrt{\frac{ NT}{2\pi  g^2 \lambda}}
=
7.25 
\sqrt{\frac{N t}{(2\pi)^3  \gbar^2 \lambda}}.
\end{equation}

\subsection{Overall pairing susceptibility}

\subsubsection{Without dynamical screening}

Combining both the semiclassical and quantum contributions, the total inverse pairing susceptibility at the transition temperature $t \rightarrow t_c$ is
\begin{equation}
	[\chi^R]^{-1}(0,\textbf{0})
	=
	-
	\frac{2 N}{W}
	+
	\frac{hN}{\pi}
	\frac{1}{\gbar^2}
	\ln
	\left(
		1 + \gbar^2 \ln \frac{1}{t_c}
	\right)
	+
	\frac{\lambda h \gbar^2 \calC}{t_c}.
\end{equation}
Since the second term has a weak temperature dependence, it just effectively renormalizes the pairing interaction strength $W$,
Eq.~(\ref{eq:Weff}). 
By combining the first two terms as $-2 N/W_{\mathsf{eff}}$, we have
\begin{equation}
	t_c
	\simeq
	\frac{\lambda h \gbar^2  \calC}{2N} W_{\sfeff},
\end{equation}
valid up to some logarithmic corrections. 
Since $\pi \lambda$ is the inverse conductance per channel, the full conductivity accounting for 
2 spin and $N$ flavor components is 
\begin{align}
	\sigma_{\sfdc}
	=
	\frac{2N}{\pi \lambda}.
\end{align}
With $h = \pi \nu_0$, $T_c = t_c \, \gammael$ and $\gammael \, \gbar^2 = g^2 / 4 \pi^2$, we recover 
Eq.~(\ref{eq:Tc})  
for $T \gg T^*$.

\subsubsection{With dynamical screening}

For $T \ll T_*$, dynamical screening becomes  important and the inverse pairing susceptibility is instead
\begin{equation}
\begin{aligned}
	[\chi^R]^{-1}(0,\textbf{0})
	&=
	-
	\frac{2 N}{W}
	+
	\frac{hN}{\pi}
	\frac{1}{\gbar^2}
	\ln
	\left(
		1 + \gbar^2 \ln \frac{1}{t_c}
	\right)
	+
	\frac{ \lambda h \gbar^2}{t_c}
	7.25 
	\sqrt{\frac{N t_c}{(2\pi)^3  \gbar^2 \lambda}}
	\\
	&=
	-
	\frac{2 N}{W}
	+
	\frac{hN}{\pi}
	\frac{1}{\gbar^2}
	\ln
	\left(
		1 + \gbar^2 \ln \frac{1}{t_c}
	\right)
	+
	7.25\, h \, 
	\sqrt{\frac{N \lambda  \gbar^2}{(2\pi)^3  t_c}}.
\end{aligned}
\end{equation}
Thus, we have (by combining the first two terms as $-2 N /W_{\mathsf{eff}}$)
\begin{equation}
t_c
=
\frac{
	\lambda \gbar^2
}{
	(2\pi)^3 
	N
	\left(\frac{2}{7.25 \, h W_{\mathsf{eff}}}\right)^2
}
=
\frac{
	\lambda \gbar^2 ( h W_{\mathsf{eff}})^2
}{
	(2\pi)^3 
	N
	\left(\frac{2}{7.25 \,}\right)^2
}
\simeq
0.05
\frac{
	\lambda \gbar^2 ( h W_{\mathsf{eff}})^2
}{
	N
}.
\end{equation}
This is Eq.~(\ref{eq:Tc}) for $T \ll T^*$.

\end{widetext}

\section{Conclusion \label{sec:Conc}}

The results of Refs.~\cite{SYK_Patel_PRB_21,SYK_Patel_PRB_2022,SYK_Guo_Patel_linearT_PRB_2022,SYK_review_Sachdev_RMP_2022,SYK_Patel_linearT_arxiv_2022,MFL_Wu_Liao_Foster_PRB_22} 
suggest that disorder plays an essential role in strange-metal physics. 
We find that interference can supply the \emph{mechanism} by which superconductivity evades suppression expected due to Planckian dissipation [Eq.~(\ref{TcS})], 
and can even be enhanced [Eq.~(\ref{eq:Tc})] in a 2D strange metal.

Our results are applicable to a broad range of quantum-critical materials where disorder is inevitable in practice,
although we do not attempt to estimate transition temperatures for particular materials here (which would depend upon details of the critical bosonic mode and disorder). 
However, we note that 
a recent experiment reported a significant increment of $T_c$ 
in Y$_5$Rh$_6$Sn$_{18}$ by atomic disorder \cite{disO_SC_expt_Maple_PRB_2020}.  
It would be interesting to further experimentally explore the effects of disorder on $T_c$ in other strongly correlated quantum materials to test our predictions. 
Another avenue for exploration would be to extend this theoretical analysis to more complicated (e.g., spatially correlated) forms of inhomogeneity \cite{Croitoru2022}.

\section{Acknowledgments}

We thank Pavel Nosov and Sri Raghu for helpful discussions. 
This work was supported by the 
Welch Foundation Grant No.~C-1809 (T.C.W.\ and M.S.F.)
and by 
the DOE (US) office of Basic Sciences Grant No.~DE-FG02-03ER46076 (P.L.).

\end{document}